\documentclass[3p]{elsarticle}
\pdfoutput=1

\usepackage{graphicx} 
\usepackage{dcolumn}
\usepackage{amssymb}
\usepackage{amsmath}
\usepackage[hidelinks]{hyperref}
\usepackage{color}
\usepackage{slashed}
\newcommand{\rmd}{\mathrm{d}}
\newcommand{\rme}{\mathrm{e}}
\newcommand{\rmi}{\mathrm{i}}
\newcommand{\tr}{\mathrm{tr}}

\newcommand{\NC}{N_{\mathrm{c}}}
\newcommand{\gmu}{\gamma^\mu}
\newcommand{\gnu}{\gamma^\nu}
\newcommand{\gp}{\gamma^+}
\newcommand{\gm}{\gamma^-}
\newcommand{\go}{\gamma^0}

\newcommand{\MDEP}{\bp,\bq,\kg}
\newcommand{\TCS}{\frac{\rmd\sigma^\gamma}{\rmd^6 K_\perp \rmd^3\eta_K}}
\newcommand{\bC}{\boldsymbol{C}}
\newcommand{\Lip}{\slashed{C}_L }
\newcommand{\Uf}{\tilde{U}}
\newcommand{\vp}{v(\bp)}

\newcommand{\ubar}{\bar{u}(\bq)}

\newcommand{\rhop}{\rho_p}
\newcommand{\rhoA}{\rho_A}
\newcommand{\qbar}{\bar{q}}
\newcommand{\pks}{\slashed{p}+\slashed{k}_\gamma}
\newcommand{\qks}{\slashed{q}+\slashed{k}_\gamma}
\newcommand{\pkprop}{\pk^2-m^2}
\newcommand{\qkprop}{\qk^2-m^2}

\newcommand{\pk}{(p+k_\gamma)}
\newcommand{\qk}{(q+k_\gamma)}
\newcommand{\qkp}{(q^++k_\gamma^+)}
\newcommand{\pkp}{(p^++k_\gamma^+)}
\newcommand{\pkm}{(p^-+k_\gamma^-)}
\newcommand{\pkpropD}{\frac{\pks-m}{\pk^2-m^2}}
\newcommand{\qkpropD}{\frac{\qks+m}{\qk^2-m^2}}

\newcommand{\calD}{\mathcal{D}}
\newcommand{\calM}{\mathcal{M}}
\newcommand{\calO}{\mathcal{O}}
\newcommand{\calP}{\mathcal{P}}
\newcommand{\calT}{\mathcal{T}}

\newcommand{\xp}{\boldsymbol{x}_\perp} 
\newcommand{\yp}{\boldsymbol{y}_\perp} 
\newcommand{\rp}{\boldsymbol{r}_\perp} 
\newcommand{\bperp}{\boldsymbol{b}_\perp} 

\newcommand{\khp}{\boldsymbol{k}_{1\perp}} 
\newcommand{\khps}{\slashed{\boldsymbol{k}}_{1\perp}} 
\newcommand{\kAp}{\boldsymbol{k}_{2\perp}} 
\newcommand{\kphp}{\boldsymbol{k}_{\gamma\perp}} 
\newcommand{\kg}{\boldsymbol{k}_\gamma} 
\newcommand{\kgp}{\boldsymbol{k}_{\gamma\perp}} 
\newcommand{\kgps}{\slashed{\boldsymbol{k}}_{\gamma\perp}} 

\newcommand{\kp}{\boldsymbol{k}_\perp}
\newcommand{\pp}{\boldsymbol{p}_\perp}
\newcommand{\qp}{\boldsymbol{q}_\perp}
\newcommand{\bk}{\boldsymbol{k}}
\newcommand{\bp}{\boldsymbol{p}}
\newcommand{\bq}{\boldsymbol{q}}
\newcommand{\kps}{\slashed{\bk}_\perp}
\newcommand{\pps}{\slashed{\bp}_\perp}
\newcommand{\qps}{\slashed{\bq}_\perp}

\newcommand{\Pp}{\boldsymbol{P}_\perp} 
\newcommand{\Pps}{\slashed{\boldsymbol{P}}_\perp} 

\begin{document}

\newpage

\begin{frontmatter}

\title{Probing gluon saturation with next-to-leading order photon production\\
       at central rapidities in proton-nucleus collisions}
\author[zagreb,tokyo]{Sanjin Beni\' c}
\author[tokyo]{Kenji Fukushima}
\author[heidelberg]{Oscar Garcia-Montero}
\author[bnl]{Raju Venugopalan}
\address[zagreb]{Physics Department, Faculty of Science,
                 University of Zagreb, Zagreb 10000, Croatia}
\address[tokyo]{Department of Physics, The University of Tokyo,
                7-3-1 Hongo, Bunkyo-ku, Tokyo 113-0033, Japan}
\address[heidelberg]{Institut f\"{u}r Theoretische Physik,
                     Universit\"{a}t Heidelberg, Philosophenweg 16,
                     69120 Heidelberg, Germany}
\address[bnl]{Physics Department, Brookhaven National Laboratory,
              Bldg.\ 510A, Upton, NY 11973, USA}
\date{\today}
\begin{abstract} 
We compute the cross section for photons emitted from sea quarks in proton-nucleus collisions at collider energies. The computation is performed within the dilute-dense kinematics of the Color Glass Condensate (CGC) effective field theory. Albeit the result obtained is formally at next-to-leading order in the CGC power counting, it provides the dominant contribution for central rapidities. We observe that the inclusive photon cross section is proportional to all-twist Wilson line correlators in the nucleus. These correlators also appear in quark-pair production; unlike the latter, photon production is insensitive to hadronization uncertainties and therefore more sensitive to  multi-parton correlations in the gluon saturation regime of QCD.
We demonstrate that $k_\perp$ and collinear factorized expressions for inclusive photon production are obtained as leading twist approximations to our result. In particular, the collinearly factorized expression is directly sensitive to the nuclear gluon distribution at small $x$. Other results of interest include the realization of the Low-Burnett-Kroll soft photon theorem in the CGC framework and a comparative study of how the photon amplitude is obtained in Lorenz and light-cone gauges. 
\end{abstract}


\end{frontmatter}

\section{Introduction}

Photons produced in proton-nucleus (p+A) collisions are powerful probes of the fundamental many-body structure of strongly interacting matter. Sufficiently energetic photons, in particular, are free of the uncertainties that attend the fragmentation of partons into hadrons. Further, the information on microscopic dynamics that is probed by photon final states complements in fundamental ways the information accessible in deeply inelastic scattering (DIS) experiments off hadrons and nuclei. A global analysis of photon production in DIS and p+A collisions, analogous to those performed for currently for parton distribution functions has therefore the potential to reveal universal features of parton dynamics that are distinct  from those that are particular to the scattering process.

The photons produced in deuteron-nucleus collisions at the ultrarelativistic energies of the Relativistic Heavy Ion Collider (RHIC) and p+A collisions at the Large Hadron Collider (LHC) are uniquely sensitive to strongly correlated states of gluons in which the gluons have the maximal occupancy allowed by QCD. The dynamics of these saturated gluon states is described by the Color Glass Condensate (CGC)~\cite{Iancu:2003xm,JalilianMarian:2005jf,Gelis:2010nm}, a classical effective theory of QCD in the high energy asymptotics (of high center of mass energies and small values of parton momentum fractions $x$) that may be applicable for significant kinematic windows at RHIC and LHC. 

A characteristic feature of the CGC is that the dynamics of saturated gluons is governed by an emergent ``saturation scale" $Q_S (x)$, which grows with increasing energy, or equivalently, decreasing $x$. Modes in hadron/nuclear wavefunctions with $k_\perp < Q_S$ are maximally occupied with an occupancy that is parametrically $1/\alpha_S$, where $\alpha_S$ is the QCD fine structure constant. In contrast, the modes $k_\perp \gg Q_S$ have small occupancies and interact dynamically as the partons of perturbative QCD (pQCD). If $Q_S$ is large relative to the intrinsic non-perturbative QCD scale, the strongly correlated dynamics of gluons can be computed using weak coupling techniques. In nuclei, the saturation scale $(Q_S^A)^2 \sim A^{1/3}$. The high energy dynamics of nuclei are therefore well suited to test the CGC description of high energy QCD.  Photon production, as noted, is a particularly sensitive probe because it is independent of the details of how partons fragment into hadrons. 

Photon production in p+A collisions was previously computed to leading order in the CGC framework~\cite{Gelis:2002ki}.  See also \cite{Dominguez:2011wm}. A derivation within the dipole formalism can be found in \cite{Kopeliovich:1998nw,Baier:2004tj}. 
Using this result, the forward prompt photon spectrum was calculated in \cite{JalilianMarian:2005zw}. Further applications include the photon-hadron \cite{JalilianMarian:2012bd,Rezaeian:2012wa} and the photon-jet correlations \cite{Rezaeian:2016szi} at forward rapidity.  

\begin{figure}
  \begin{center}
  \includegraphics[scale = 1.25]{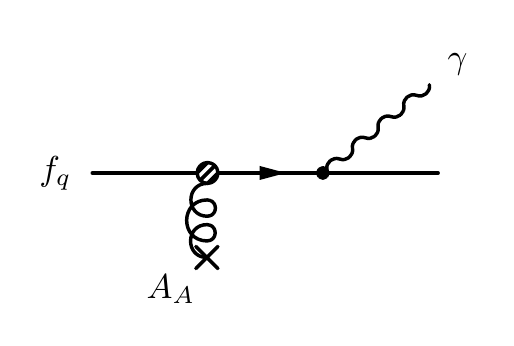}
  \end{center}
  \caption{Leading  order process for prompt photon production in proton-nucleus collisions.  The diagram describes the bremsstrahlung of a photon from a valence quark after multiple scattering off the classical gluon field in the nucleus.}
  \label{fig:LO}
\end{figure}

The power counting for proton-nucleus collisions corresponds to a ``dilute-dense" limit, where contributions to lowest order in $Q_S^p/k_{1\perp} \ll 1$ (where $Q_S^p$ is the saturation scale in the projectile proton) are preserved along with all order terms in $Q_S^A/k_{1\perp}$. Here $k_{1\perp}$, $k_{2\perp}$ respectively correspond to the momentum exchange from the proton and the nucleus to the final state of interest. For photon production, the leading term in this dilute-dense power counting corresponds to order $O(\alpha)$, where $\alpha_e$ is the QED fine structure constant.  This leading order  (LO) contribution is illustrated in Fig.~\ref{fig:LO}. The quark line in this figure corresponds to a valence quark in the wavefunction of the projectile proton. In the high occupancy regime of $k_\perp\leq Q_S^A$, there is no $\alpha_S$ dependence at LO because the $\alpha_S$ factor in the cross section arising from the coupling of a gluon to the valence quark is compensated by the $1/\alpha_S$ occupancy of these gluons in the target. The $O(\alpha)$ dependence must be understood as being accompanied by the valence quark distribution $f_q$ in the proton. 

\begin{figure}
  \begin{center}
  \includegraphics[width=\textwidth]{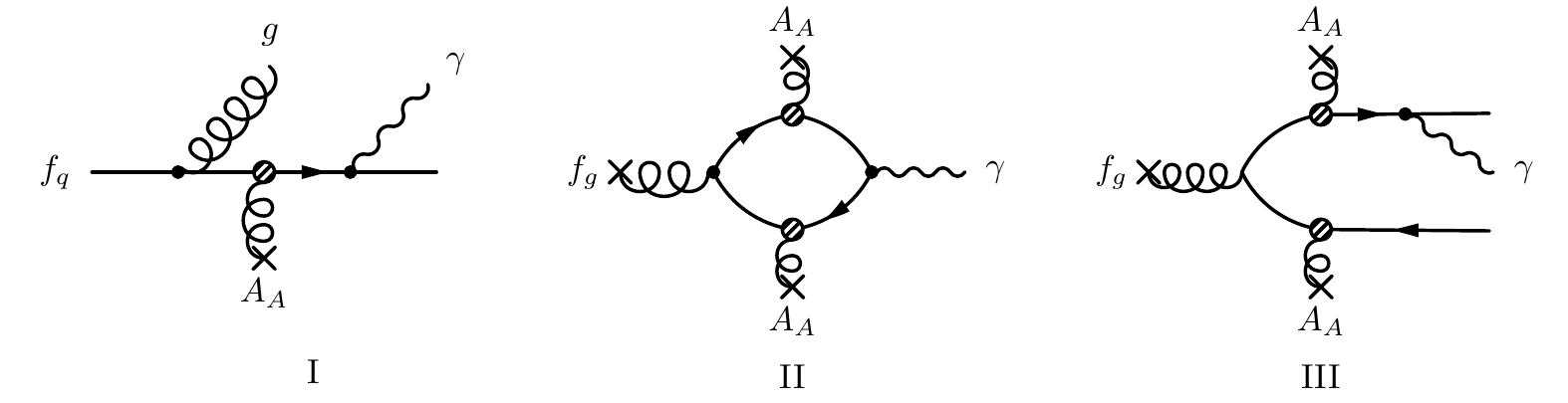}
  \end{center}
  \caption{Next-to-leading order processes for prompt photon production in proton-nucleus collisions.  The class I diagram describes the bremsstrahlung of a photon from a valence quark. The next two diagrams correspond to a gluon from the proton splitting into a $q\qbar$ that  annihilates into a photon final state (class II), or emits a photon either before or after rescattering off the nucleus (class III). As described in the text, the class I diagram, upon evolution, is the 
same order as the diagram in Fig.~\ref{fig:LO}.}
  \label{fig:classes}
\end{figure}

At next-to-leading order (NLO) $O(\alpha_e\alpha_S)$, there are a number of contributions which can be classified into the three classes shown in Fig.~\ref{fig:classes}. The leftmost diagram (class I) corresponds to a gluon emitted from the incoming valence quark. For inclusive photon production, one has to integrate over the phase space corresponding to the emitted gluon.  Another NLO contribution to inclusive photon production in this diagrammatic class (not shown) arises from the interference between the LO contribution in the amplitude and a  contribution in the complex conjugate amplitude corresponding to the virtual emission and absorption of a gluon by the valence quark. Since both of these diagrams are of a bremsstrahlung type, the divergence structure is inherited from NLO quark production, investigated in detail in Ref.~\cite{Chirilli:2012jd}. The resulting expression gives
logarithms that are sensitive to the transverse momentum and $x$ of the gluon as well as finite pieces.
These diagrams contribute to the double-log DGLAP renormalization group (RG) evolution of the valence quark distribution\footnote{The dominant contribution comes from the large phase space in transverse momentum $\alpha_S\ln(k_\perp)\sim 1$; since valence quarks are predominantly localized at $x\sim 1$, the logarithms in $x$ are sub-dominant, as is the case for DGLAP evolution.}.
The finite pieces can be absorbed in the definition of the quark distribution function by appropriate choice of factorization scale and by choice of factorization scheme.
The NLO contributions of class I are therefore actually of $O(\alpha_e)$ if the bare valence quark distribution in the proton is replaced by the RG evolved quark distribution. 

The class II 
diagram in Fig.~\ref{fig:classes} was computed recently~\cite{Benic:2016yqt}. In this case, since 
the quark-antiquark pair are emitted from the gluon prior to their subsequent annihilation into the photon. 
At small-$x$ kinematics, this
diagram is of order $O(\alpha_e\alpha_S)$ with the cross section for this contribution  accompanied by a factor $f_g$ corresponding to the gluon distribution in the proton. 
The emission of the initial gluon from the valence quark line is of order $\alpha_S \log 1/x \sim 1$ at small-$x$ where it does not contribute to the power counting.
While this is an NLO diagram, it can in principle provide a much larger contribution to photon production. This is because the gluon distribution grows rapidly while the valence quark distribution decreases at small $x$, giving  $f_g \gg f_q$. Thus for photon production in the small $x$ kinematics of the projectile proton, such NLO contributions can dominate significantly. The kinematics of the class II diagram is relevant for inclusive photon production at central rapidities. Because of pair annihilation, the transverse momentum of the photon for this diagram is strongly constrained to be dominated by momenta around $k_{\gamma\perp}\sim Q_S^A$. Thus this contribution is in principle very sensitive to the saturation scale of the nucleus.

However as also noted in Ref.~\cite{Benic:2016yqt}, there is a further class of NLO processes, class III in Fig.~\ref{fig:classes}, that contribute significantly to photon production in p+A collisions. Firstly, like the class II process, this NLO contribution comes accompanied by a factor $f_g$ which, as noted, will overwhelm the LO contribution at small $x$. Secondly there are some features of the class III computation that are qualitatively different from those of class II. Unlike the latter, photon production, while sensitive to $Q_S^A$, is dominated by soft momenta with $k_\perp < Q_S^A$. Similarly, since photon production is not as kinematically constrained for class III diagrams relative to class II diagrams, it will also dominate at large $k_{\gamma\perp} > Q_S^A$. In particular, class III diagrams will match the contribution from leading twist pQCD at high $k_{\gamma\perp}$, while the class II contribution is proportional to a higher twist four point correlator even at large $k_{\gamma\perp}$. The sum of class II and class III diagrams constitute the relevant NLO contribution to inclusive photon production in the CGC framework. 

In this paper, we will compute the class III NLO diagrams for photon production in p+A collisions in the CGC framework\footnote{Several of the results presented here were first obtained as a part of the Masters's thesis of one of the authors (Garcia-Montero) at the University of Heidelberg~\cite{Garcia-Montero}.}. We will perform the computation first in 
Lorenz gauge $\partial_\mu A^\mu=0$ gauge, and subsequently in light-cone gauge $A^+=0$. In addition to being an independent non-trivial check of our results, the intermediate steps are interesting and realized differently in the two gauges. 

The paper is organized as follows. In Section \ref{sec:AL}, after a preliminary discussion of dilute-dense collisions in the CGC framework, we will outline the derivation of the amplitude for the inclusive production of a $q\bar{q}\gamma$ final state in Lorenz gauge. Our work closely follows the previous derivation in this gauge of the amplitude for gluon production~\cite{Blaizot:2004wu} and quark-antiquark pair production~\cite{Blaizot:2004wv} in proton-nucleus collisions.  As for the case of the pair production amplitude considered previously, we show that contributions from so-called ``singular" terms, wherein the photon is produced from within the target, are exactly canceled by terms that one can identify as gauge artifacts in regular terms. The latter are contributions where the quark-antiquark pair (and the photon) are produced either before or after the scattering of gluons from the target off the projectile.

In Section \ref{sec:CS}, we compute the cross section for inclusive photon production. 
For readers interested in the central result of this work, the key expression is given in Eq.~(\ref{eq:Full4}).
As in the case of quark-antiquark production~\cite{Blaizot:2004wv}, the cross section factorizes into the product of the unintegrated gluon distribution in the projectile times the sum of terms corresponding to the gluon distribution in the target,  and quark-antiquark-gluon and quark-antiquark-quark-antiquark light-like Wilson line correlators. These correlators contain non-trivial information on many-body gluon and sea quark correlations that are of all twist order in conventional pQCD language. In Sec.~\ref{sec:KF} we demonstrate that for $Q_S^A \ll k_{2\perp}$, the cross section can be expressed as a $k_\perp$ factorized product of unintegrated gluon distributions in the projectile and target \cite{Baranov:2007np,Motyka:2016lta}.
We show explicitly that the corresponding leading twist  $k_\perp$ factorized amplitude agrees exactly with the expression derived recently by Motyka, Sadzikowski and Stebel~\cite{Motyka:2016lta} in the context of $Z^0$ boson hadroproduction.  Taking the limit $k_{1\perp}$, $k_{2\perp}\to 0$, one recovers the formal structure of the leading collinear factorization contribution to inclusive photon production from gluon-gluon scattering~\cite{Gordon:1993qc}. This demonstrates the unique sensitivity of inclusive photon production to the nuclear gluon distribution function at small x. Unfortunately, the detailed analytical comparison of the two expressions is cumbersome; a numerical comparison is left for future work. In Sec.~\ref{sec:SO}, we summarize our results and provide a detailed outline of the necessary ingredients for the numerical computation of inclusive photon production in proton-nucleus collisions. We will leave this numerical computation for a subsequent publication. 

In \ref{app:LC} we will compute the amplitude for inclusive photon production in light-cone gauge $A^+=0$. This computation follows a previous computation of gluon production in this gauge~\cite{Gelis:2005pt}. Unlike gluon production, where the amplitudes in Lorenz gauge and light-cone gauge differ (albeit the cross sections of course agree), in this case, the amplitudes in the two gauges agree exactly. 
Some properties of the amplitude are outlined in \ref{app:Properties}. In addition to demonstrating that this amplitude satisfies a Ward identitiy, we show in particular that the well known Low-Burnett-Kroll theorem~\cite{Low:1958sn,Burnett:1967km,Bell:1969yw} is satisfied. This corresponds, in the soft photon limit, to the factorization of the amplitude into the amplitude for quark-antiquark production times a contribution determined by the Lorentz structure of the photon.

\section{Amplitude for inclusive $q\bar{q}\gamma$ production in the Lorenz  gauge}
\label{sec:AL}

In the CGC effective theory, the gluon dynamics with high occupancy at
small $x$ are described by the classical Yang-Mills equations.  
For collisions of the proton moving in the positive $z$
direction and the nucleus moving in the negative $z$ direction at the
speed of light, the Yang-Mills equations are
\begin{equation}
  [D_\mu, F^{\mu\nu}](x) = g \delta^{\nu +}\delta(x^-)\rhop(\xp)
    + g \delta^{\nu - }\delta(x^+)\rhoA(\xp)\,,
\label{eq:YME}
\end{equation} 
where $\rhop$ and $\rhoA$ correspond to the static and stochastic
color charge density matrices at large $x$ in the proton and the
nucleus, respectively.  This framework is feasible at a given impact
parameter for three different regimes of color charge densities
depending on the energy of the collision and the kinematic regime of interest.
These correspond to the dilute-dilute ($\rhop/\khp^2 \ll 1$,
$\rhoA/\kAp^2 \ll 1$), the dilute-dense ($\rhop/\khp^2 \ll 1$,
$\rhoA/\kAp^2 \sim 1$), and the dense-dense ($\rhop/\khp^2 \sim 1$,
$\rhoA/\kAp^2 \sim 1$) regimes.  To compute inclusive cross sections,
the modulus squared of the computed amplitude must be averaged over
the color sources, thereby replacing
$\rhop,\rhoA \rightarrow Q_S^p,Q_S^A$ in the
power counting.  Subsequent to this averaging, the dilute-dilute limit in the CGC
effective theory can be matched to the leading twist frameworks of
pQCD at small $x$. Further, the dilute-dense limit corresponds to the dominance
of leading twist contributions on the proton side with all twist
contributions on the nuclear side, and the dense-dense limit includes
all twist contributions from both the proton and the nucleus to the
scattering process of interest.

At the collider energies currently accessible in p+A collisions, the
appropriate dynamics is that of the dilute-dense limit.  The nucleus
is dense because there is an enhancement of the order of $A^{1/3}$ in the number of 
color sources even at large $x$. These sources can all radiate soft gluons
that further function as color sources for even softer gluons.  In
the proton, on the other hand, there are $O(1)$ color sources, and it
is only at very large rapidities relative to the proton fragmentation
region that the density of color sources becomes 
$\rhop/\khp^2\sim 1$.  Alternately, the dense-dense regime can be
attained in very rare high multiplicity events where the proton
fluctuates into sources of color charge already at large $x$. 

In this work, we will be interested in photon production in p+A
collisions at collider energies for the small $x$ kinematics, and in
event classes where the dilute-dense asymptotics is appropriate and
gives the dominant contribution.  Analytic computations are feasible
then, and explicit expressions can be derived.  In contrast, the
dense-dense limit is not analytically tractable even for inclusive
gluon production since there is no small parameter to expand in. Results can be obtained only through numerical simulations of the Yang-Mills equations~\cite{Krasnitz:1998ns,Krasnitz:2001qu,Lappi:2003bi}.

\subsection{Structure of the gluon field and setup of the amplitude computation}

In what follows, we work entirely in the Lorenz covariant gauge $\partial_\mu A^\mu = 0$. The outline of the derivation in light-cone gauge 
is given in \ref{app:LC}.  For a dilute-dense
system with $\rhop/\khp^2\ll 1$ and $\rhoA/\kAp^2\sim 1$, the
analytical solution of Eq.~\eqref{eq:YME} in the Lorenz gauge can be
expressed in momentum space
as~\cite{Blaizot:2004wu}
\begin{equation}
  A^\mu(q) = A_p^\mu(q)
    + \frac{\rmi g}{q^2+\rmi q^+\epsilon}\int_{\khp}\int_{\xp}
    \rme^{\rmi(\qp-\khp)\cdot\xp} \Bigl\{ C^\mu_U(q,\khp)[U(\xp)-1]
    + C^\mu_V(q)[V(\xp)-1] \Bigr\} \frac{\rhop(\khp)}{\khp^2}\,,
\label{eq:field}
\end{equation}
where $A_\mu(q) = A_\mu^a(q) T^a$ and $T^a$ are the generators of SU($\NC$) in the adjoint representation.
We will also use a shorthand notation,
$\int_{\kp}\equiv\int\frac{\rmd^2\kp}{(2\pi)^2}$ and
$\int_{\xp}\equiv\int\rmd^2\xp$, hereafter.
In detail, the ingredients going into Eq.~\eqref{eq:field} are as follows.
The first term, $A^\mu_p(q)$, represents the gluon field of the proton alone
\begin{equation}
  A_p^\mu(x) = - g\delta^{\mu +}\delta(x^-)\frac{1}{\nabla_\perp^2}\rhop(\xp)\,.
\label{eq:pA}
\end{equation}

The vectors $C^\mu_U$ and $C^\mu_V$ are abbreviated forms for the
momentum dependence of the integral, with $\khp$ being the momentum
exchanged from the proton, while $\kAp\equiv\qp-\khp$ exchanged from the
nucleus.  The explicit forms of $C^\mu_U$ and $C^\mu_V$ are
\begin{equation}
  \begin{split}
 & C_U^+(q,\khp) \equiv -\frac{\khp^2}{q^-+\rmi\epsilon}\,, \qquad
   C_U^-(q,\khp) \equiv \frac{(\qp-\khp)^2-\qp^2}{q^+}\,, \qquad
   \bC_U^\perp(q,\khp) \equiv -2\khp\,, \\
 & C_V^+(q) \equiv 2q^+\,, \qquad
   C_V^-(q) \equiv 2\frac{\qp^2}{q^+}-2q^-\,, \qquad
   \bC_V^\perp(q) \equiv 2\qp\,.
  \end{split}
\label{eq:Lip1}
\end{equation}
These two functions are related to the well-known Lipatov effective
vertex \cite{Lipatov:1976zz,Kuraev:1977fs,Balitsky:1978ic} via the
simple relation, $C^\mu_L = C^\mu_U +C^\mu_V/2$.  This effective vertex is a gauge
covariant expression that efficiently combines the various
contributions to glue-glue scattering in the Regge-Gribov limit of
QCD.

In Eq.~\eqref{eq:field}, the Wilson lines $U$ and $V$ account for
the modification of the gluon field of the proton due to multiple gluon scatterings in the nucleus
and can be expressed, for an arbitrary light-like path as
\begin{align}
  U(a,b;\xp) &= \calP_+ \exp\biggl[ \rmi g\int^a_b \rmd z^+
    A^-_A(z^+,\xp)\cdot T\biggr]\,,
\label{eq:Wilson}\\
  V(a,b;\xp) &= \calP_+ \exp\biggl[ \frac{\rmi g}{2} \int^a_b
    \rmd z^+ A^-_A(z^+,\xp)\cdot T\biggr]\,,
\end{align}
where $A^-_A(z^+,\xp)$ represents the gluon field alone. This field can be expressed similarly to $A_p^\mu$ above as 
\begin{equation}
 A_A^\mu(x) = - g\delta^{\mu -}\delta(x^+)\frac{1}{\nabla_\perp^2}\rhoA(\xp)\,.
\label{eq:AA}
\end{equation}
In the above expressions for the Wilson lines $\calP_+$ denotes
path-ordering in the $x^+$ direction.

While $U$ is the standard Wilson line in the adjoint
representation of SU($\NC$),  $V$ is an unusual form of the Wilson line that turns out to be a gauge artifact of the gluon production amplitude in Lorenz gauge.  This can be deduced from the fact that the $V$ dependent terms in the amplitude do not appear in the cross sections nor in calculations
in other gauges~\cite{Gelis:2005pt,Dumitru:2001ux}. While one expects therefore that $V$ should drop out at the level of
the cross section for photon production, we will show that they will not appear in our final expressions for the amplitude. The same observation was made 
previously for the amplitude for $q\bar{q}$ production in Lorenz gauge.  
We will henceforth use the
following shorthand notation for the complete Wilson lines,
\begin{equation}
  U(\xp) \equiv U(\infty,-\infty;\xp)\,,\qquad
  V(\xp) \equiv V(\infty,-\infty;\xp)\,.
\end{equation} 

Before we proceed, it is instructive to discuss the underlying
structure of Eq.~\eqref{eq:field} in coordinate space.  This can be
decomposed by splitting Eq.~\eqref{eq:field} as $A^\mu = A_R^\mu+A_S^\mu$
where $A_R^\mu$ is called a regular term and $A_S^\mu$ a singular
term.
The former corresponds to the emission of the gluon from the proton
before interacting with the highly Lorentz contracted shock wave of
gluons that comprises the nuclear target, while the latter corresponds to gluon production from within the shock wave. The latter term is 
proportional to the Lorentz contracted width of the nucleus $\propto \delta(x^+)$.  
Because
$2q^-=q^2/q^+ + \qp^2/q^+$, we can split $C_V^-(q)$, the only
term with $q^-$ dependence, into a regular part $C^\mu_{V,\text{reg}}(q)$
and a ``singular" part
\begin{equation}
  C_V^\mu(q) = C^\mu_{V,\text{reg}}(q) - \delta^{\mu-}\frac{q^2}{q^+}\,.
\label{eq:reg}
\end{equation}
We observe that the singular field appears from the second term in
Eq.~\eqref{eq:reg} because the $q^2$ term in Eq. \eqref{eq:reg} cancels the $1/q^2$ pole in Eq.~\eqref{eq:field}.  Clearly, the $q^-$ integration leads to
$\delta(x^+)$, but we need to be cautious about the infinitesimal
longitudinal extension in $V(\xp)$.  After careful treatment, its
Fourier transformed representation in coordinate space can be expressed as 
\begin{equation}
  A_S^\mu(x) = \frac{g^2}{2} A_A^\mu(x) V(x^+,-\infty;\xp)
    \theta(x^-)\frac{1}{\nabla^2_\perp}\rhop(\xp)\,.
\end{equation}

In addition to decomposing the gauge field, we will need one more
ingredient for our computation of the class III amplitude for photon
production.  This is the effective vertex corresponding to self-energy
insertions in the time-ordered quark propagator arising from multiple
insertions of the  nuclear gluon background field: it is given 
by~\cite{Blaizot:2004wv}
\begin{equation}
  \calT(k,p) = \begin{cases}
    \displaystyle
    2\pi\delta(k^+)\gp\int_{\xp}
      \rme^{\rmi\kp\cdot\xp}\bigl[ \Uf(\xp)-1 \bigr] & (p^+>0)\,,\\
    \displaystyle
    -2\pi\delta(k^+)\gp\int_{\xp}
      \rme^{\rmi\kp\cdot\xp}\bigl[ \Uf^\dag(\xp)-1 \bigr] & (p^+<0)\,,
  \end{cases}
\label{eq:fundvert}
\end{equation}
where $p$ is the quark momentum and $k$ is the momentum transfer from
the multiple gluon ``kicks'' to the quark.  The Wilson line in the
fundamental representation $\Uf$ is defined as in
Eq.~\eqref{eq:Wilson} with the SU($\NC$) adjoint generators $T^a$
replaced by the generators $t^a$ of the fundamental representation. Explicitly, it is written as~\cite{McLerran:1998nk}
\begin{equation}
  \Uf(\xp) = \calP_+ \exp\biggl[ -\rmi g^2\int_{-\infty}^\infty
    \rmd z^+ \frac{1}{\nabla_\perp^2} \rhoA(z^+,\xp)\cdot t\biggr]\,.
\label{eq:Utilde}
\end{equation}
The effective vertex appears in the quark propagator and $\Uf$ and $\Uf^\dag$
sum over all the gluon insertions to the quark and the antiquark
respectively.  The momentum transfer from the nucleus $k$ should be
integrated over.  Since the effective vertex does not change the order
of the diagrams parametrically by any power of $g$, processes where the emitted photon is sandwiched between two such effective vertices are possible.
We will see in the following that this is not the case;  the
kinematics of the process will constrain the number and configuration
of these effective vertices.

\begin{table}
  \begin{center}
  \begin{tabular}{clp{1em}cl}
    \hline
    $k_1$: & 4-momentum exchanged from the proton &&
    $k_2$: & 4-momentum exchanged from the nucleus \\
    $k_\gamma$: & photon 4-momentum &&
    $k$: & 4-momentum from $\Uf$ \\
    $q$: & quark 4-momentum && $p$: & antiquark 4-momentum \\
    $P$: & \multicolumn{4}{l}{4-momentum of the final state $P=k_1+k_2=p+q+k_\gamma$} \\ 
         & \multicolumn{4}{l}{(4-momentum from $\Uf^\dag$: $k_2-k=P-k_1-k$)} \\
    \hline
  \end{tabular}
  \end{center}
  \caption{Summary of momentum notations used in the text.}
\label{tab:not}
\end{table}
From the above discussion, we can deduce that for the amplitude for photon
production in Fig.~\ref{fig:classes} there are fourteen non-vanishing
contributions.  We will separate them in two groups of diagrams:
\begin{enumerate}
  \item the regular terms, in which the $q\qbar$ pair is created from
    the regular field $A^\mu_R$.
  \item the singular terms, wherein the pair is spawned from $A^\mu_S$.
\end{enumerate}
In the following computation, we shall include only one insertion of
$A_{R/S}^\mu$ on the quark propagator, regular or singular, to stay consistently at first
order in the proton source $\rhop$.  As noted, the number of gluon insertions from the nucleus 
onto the quark and antiquark lines from the nucleus can, however, in
our dilute-dense power counting, be as many as the kinematics permit.
The amplitude can be 
decomposed into the external polarization vector of the photon and
an amplitude vector, with the results in the following subsections expressed in terms of the amplitude vector defined as
\begin{equation}
  \calM_\lambda(\MDEP) \equiv \epsilon_\mu^\ast(\kg,\lambda)\calM^\mu(\MDEP)\,,
\end{equation}
where $\bq$, $\bp$, and $\kg$ are the quark, the antiquark, and the
photon external three momenta, respectively, and $\lambda$ is the photon polarization. We summarize in Table \ref{tab:not} the momenta notations that will be used in the following calculations.

\subsection{Regular contributions to the amplitude}

\begin{figure}
  \begin{center}
  \includegraphics[scale=1.0]{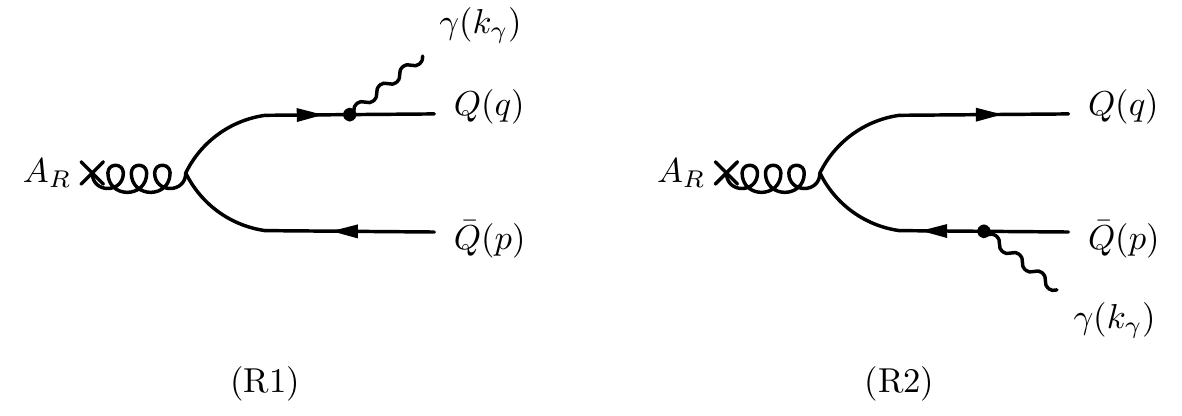}
  \end{center}
  \caption{Brehmstrahlung diagrams without gluon insertions in the
    quark and antiquark lines.}
\label{fig:D0}
\end{figure}

Following the above stated classification, we will proceed to find
the regular diagrams which have no fundamental Wilson lines first.
For this case, there are two
diagrams, shown on Fig.~\ref{fig:D0},
with exactly
one insertion of the proton source, in the form of the regular field $A_R$.
The two diagrams represent the scattering of a gluon off the target and the
resulting creation of a $q\qbar$ pair.  The photon is then emitted
from the quark or antiquark line.  Using standard Feynman rules, the
vector amplitude for the diagram (R1) (denoted as $\calM^\mu_{R1}$) is
\begin{equation}
  \calM_{R1}^\mu(\MDEP) = \ubar(-\rmi q_f e \gamma^\mu) S_0(q+k_\gamma)(-\rmi g
    \slashed{A}(P)\cdot t)\, \vp\,,
\label{eq:MA1}
\end{equation}
where $P$ is the total external 4-momentum 
\begin{equation}
P \equiv p+q+k_\gamma\, ,
\end{equation}
and the
quark lines are given in this calculation by the vacuum time-ordered
fermion propagator
\begin{equation}
S_0(p) \equiv \rmi \frac{\slashed{p} + m}{p^2 - m^2 + i\epsilon}\,.
\end{equation}

The proton field $A_p^\mu$ does not contribute to this diagram.
It contains the delta function $\delta(p^- + q^- + k_\gamma^-)$
which cannot be satisfied if the quark, antiquark, and photon are
on-shell, as  $p^-, q^-, k_\gamma^- >0$.  Dropping the
$A_p^\mu$ term, we are left with the rest of $A_R^\mu$, which, for the amplitude $\calM_{R1}^\mu$, gives,
\begin{equation}
  \begin{split}
    \calM_{R 1}^\mu(\MDEP) &= \frac{q_f e g^2}{P^2}\int_{\khp}\int_{\xp}
    \frac{\rhop^a(\khp)}{\khp^2}\,\rme^{\rmi(\Pp-\khp)\cdot\xp}\,
    \ubar\gmu \qkpropD \\
    &\qquad \times\Bigl\{ \bigl[U(\xp)-1\bigr]^{ba}\slashed{C}_U(P,\khp)
    +\bigl[V(\xp)-1\bigr]^{ba} \slashed{C}_{V,\text{reg}}(P)
    \Bigr\}t^b \vp\,.
  \end{split}
\label{eq:MR1}
\end{equation}
Following the same procedure as the one for (R1), one finds the
amplitude contribution (R2) for the photon emitted from the antiquark
to be
\begin{equation}
  \begin{split}
    \calM_{R2}^\mu(\MDEP) &= -\frac{q_f e g^2}{P^2}\int_{\khp}\int_{\xp}
    \frac{\rhop^a(\khp)}{\khp^2}\,\rme^{\rmi(\Pp-\khp)\cdot\xp}
    \ubar\Bigl\{ \bigl[U(\xp)-1\bigr]^{ba}\slashed{C}_U(P,\khp) \\
    &\qquad + \bigl[V(\xp)-1\bigr]^{ba} \slashed{C}_{V,\text{reg}}(P)
    \Bigr\}\pkpropD\gmu t^b\vp\,.
  \end{split}
\label{eq:MR2}
\end{equation}
It will be shown in the next subsection that the singular
contributions will cancel the terms with the Wilson line $V$--the final
expression has no dependence on $V$. 

\begin{figure}
  \begin{center}
  \includegraphics[width=\textwidth]{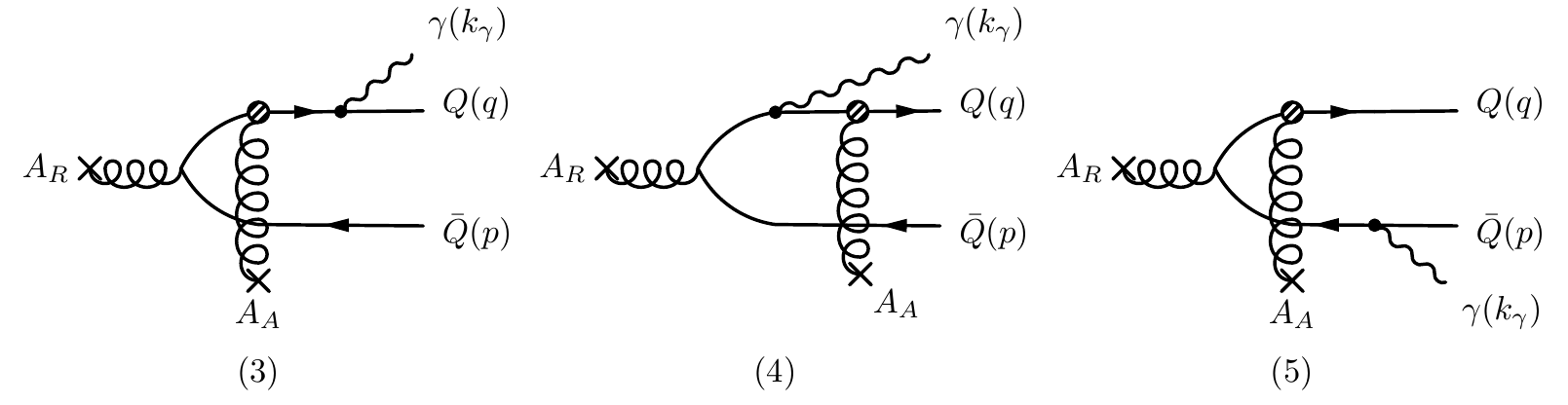}
  \includegraphics[width=\textwidth]{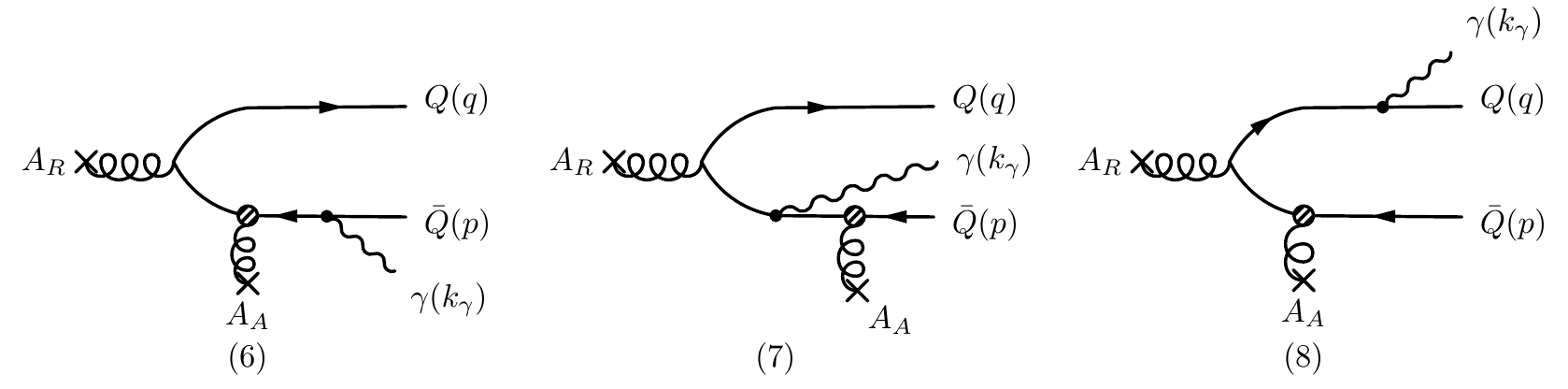}
  \end{center}
  \caption{Contributions for the amplitude with at most one Wilson
    line in the fundamental representation, either $\Uf$ or $\Uf^\dag$.}
\label{fig:Reg1}
\end{figure}

Diagrams with one insertion of the effective vertex on the quark propagator can have one Wilson line $\Uf$ or
$\Uf^\dag$ in the fundamental representation, as shown in
Fig.~\ref{fig:Reg1}, for the quark [diagrams (3)-(5)] and likewise for
the antiquark [diagrams (6)-(8)].  In the following steps we will treat them separately for
convenience.  As in the case of the amplitudes
(1) and (2), the amplitude (3) in Fig.~\ref{fig:Reg1} will have a
regular field insertion.  However, now in addition we must insert the
effective nuclear vertex \eqref{eq:fundvert} for the multiple gluon scatterings.  We
should integrate over nuclear momentum transfer $k_2$ to obtain,
\begin{equation}
  \calM_3^\mu(\MDEP) = \int\frac{\rmd^4 k_2}{(2\pi)^4}
    \ubar(-\rmi q_f e\gamma^\mu) S_0(q+k_\gamma)
    \calT(k_2,q+k_\gamma) S_0(q+k_\gamma-k_2)(-\rmi g
    \slashed{A}_R(P-k_2)\cdot t)\,\vp\,.
\label{eq:m3lore}
\end{equation}
We then integrate over $k_2^+$ and $k_2^-$.  This integration is
trivial for $k_2^+$ since $\calT(k_2,q+k_\gamma)$ contains $\delta(k_2^+)$.  
Only the proton field part $A_p^\mu$ of the regular field gives a finite contribution.
In this part,
the $k_2^-$ integration is also trivial
because $A_p^\mu$ contains $\delta(P^--k_2^-)$.
We shall now demonstrate that the integration over the remaining part of $A_R^\mu$ vanishes by the residue theorem.
The singularities in $k_2^-$ of the regular field and the quark propagator
$S_0(q+k_\gamma-k_2)$, respectively, are
\begin{equation}
  \begin{split}
    & k_2^- = P^- - \frac{(\Pp-\kAp)^2}{2P^+}
        + \frac{\rmi\varepsilon}{2P^+}\,,\qquad
      k_2^- = P^- + \rmi\varepsilon\,,\\
    & k_2^- = q^- + k_\gamma^-
        - \frac{m^2+(\qp+\kgp-\kAp)^2}{2(q^++k_\gamma^+)}
        + \frac{\rmi\varepsilon}{2(q^++k_\gamma^+)}\,,
  \end{split}
\end{equation}
where the first pole comes from the prefactor and the second from $C_U^+$ in $A^\mu_R$.
Since all the three poles are above the real $k_2^-$ axis, the $k_2^-$ integral vanishes by the residue theorem.
This result has the
clean physical interpretation that the quark-antiquark pair is first
created in the proton and subsequently the quark scatters off the gluons in the
nucleus.  
The calculation of the diagrams (4) and (5) is analogous.
We can collect the results for the amplitude vector in a compact form as
\begin{equation}
  \calM_\beta^\mu(\MDEP) = -q_f e g^2 \int_{\khp}\int_{\xp}
    \frac{\rhop^a(\khp)}{\khp^2}\,\rme^{\rmi(\Pp-\khp)\cdot\xp}
    \ubar R_\beta^\mu(\khp)\bigl[\Uf(\xp)-1\bigr]t^a\vp \,,
\label{eq:OIQ}
\end{equation}
where the integration variable was changed from $k_2$ to $k_1=P-k_2$ and $\beta\in 3,4,5$.  The symbol $R_\beta^\mu(\khp)$ is a shorthand
notation for the Dirac structure
\begin{equation}
  \begin{split}
    R_3^\mu(\khp) &\equiv \gmu \qkpropD\gp
    \frac{\khps-\pps+m}{2\qkp p^- + M^2(\khp-\pp)}\gm \,,\\
    R_4^\mu(\khp) &\equiv -\gp 
    \frac{\slashed{q}+\khps-\Pps+m}{2q^+\pkm+M^2(\khp-\pp-\kgp)}
    \gmu\frac{\khps-\slashed{p}+m}{2\qkp p^- + M^2(\khp-\pp)}\gm \,,\\
  R_5^\mu(\khp) &\equiv \gp
    \frac{\pps+\kgps-\khps-m}{2q^+\pkm+M^2(\khp-\pp-\kgp)}
    \gm \pkpropD\gmu\,,
  \end{split}
\label{eq:R345}
\end{equation}
where we define $\pps \equiv p_i \gamma^i$ and $M^2(\pp)\equiv \pp^2 + m^2$ as the transverse mass.
The reductions to transverse Dirac matrices in the numerators comes from using $k_1^+ = P^+$, $k_1^- =0$ and also the identities $(\gamma^+)^2 = 0$, $(\gamma^-)^2 = 0$.

The next three contributions to the photon amplitude, represented in
diagrams (6)--(8) of Fig.~\ref{fig:Reg1}, can be found in the same
fashion, and as the former, can be shown to have vanishing pole
integrations.  The amplitudes for these processes are
\begin{equation}
  \calM_\beta^\mu(\MDEP) = -q_f e g^2\int_{\khp}\int_{\xp}
    \frac{\rhop^a(\khp)}{\khp^2}\,\rme^{\rmi(\Pp-\khp)\cdot\xp}
    \ubar R_\beta^\mu(\khp)t^a\bigl[\Uf^\dag(\xp)-1\bigr]\vp\,,
\end{equation}
where $\beta\in 6,7,8$ corresponds to the respective diagrams in
Fig.~\ref{fig:Reg1} and the corresponding
Dirac structures
\begin{equation}
  \begin{split}
  R_6^\mu(\khp) &\equiv \gm\frac{\qps-\khps+m}
    {2\pkp q^- + M^2(\khp-\qp)}\gp \pkpropD\gmu\,,\\
  R_7^\mu(\khp) &\equiv -\gm\frac{\slashed{q}-\khps+m}
    {2\pkp q^-+ M^2(\khp-\qp)}\gmu\frac{\slashed{p}+\khps-\Pps-m}
    {2p^+(q^-+k_\gamma^-) + M^2(\qp+\kgp-\khp)}\gp\,,\\
  R_8^\mu(\khp) &\equiv \gmu \qkpropD\gm
    \frac{\khps-\qps-\kgps-m}{2p^+(q^-+k_\gamma^-) + M^2(\qp+\kgp-\khp)}\gp\,.
  \end{split}
\label{eq:R678}
\end{equation}
To tidy up our notation, we will express the sum of the contributions
(3)--(8) in Fig.~\ref{fig:Reg1} as
\begin{equation}
  \begin{split}
    \sum_{\beta=3}^8 \calM_\beta^\mu(\MDEP) &= -q_f e g^2
      \int_{\kp\khp}\int_{\xp\yp}\frac{\rhop^a(\khp)}{\khp^2}\,
      \rme^{\rmi\kp\cdot\xp+ \rmi(\Pp-\kp-\khp)\cdot\yp} \\
    &\quad\times \ubar \Bigl\{T^\mu_q(\khp)\bigl[\Uf(\xp)-1\bigr]
      t^a + T^\mu_{\qbar}(\khp)t^a\bigl[\Uf^\dag(\yp)-1\bigr]
      \Bigr\}\vp\,,
  \end{split}
\label{eq:qq}
\end{equation}
where the total Dirac structure
is combined as
\begin{equation}
T^\mu_q(\khp) \equiv \sum_{\beta=3}^5 R^\mu_\beta(\khp) \,, \qquad
T^\mu_{\qbar}(\khp) \equiv \sum_{\beta=6}^8 R^\mu_\beta(\khp)\,.
\label{eq:TqTqbar}
\end{equation}
In the first term in \eqref{eq:qq}, we introduced a dummy integration over $\yp$ and $\kp$. In the second term in \eqref{eq:qq}, we renamed $\xp\to\yp$ and further, introduced a dummy integration over $\xp$ and $\kp$.

\begin{figure}
  \begin{center}
  \includegraphics[scale=1.0]{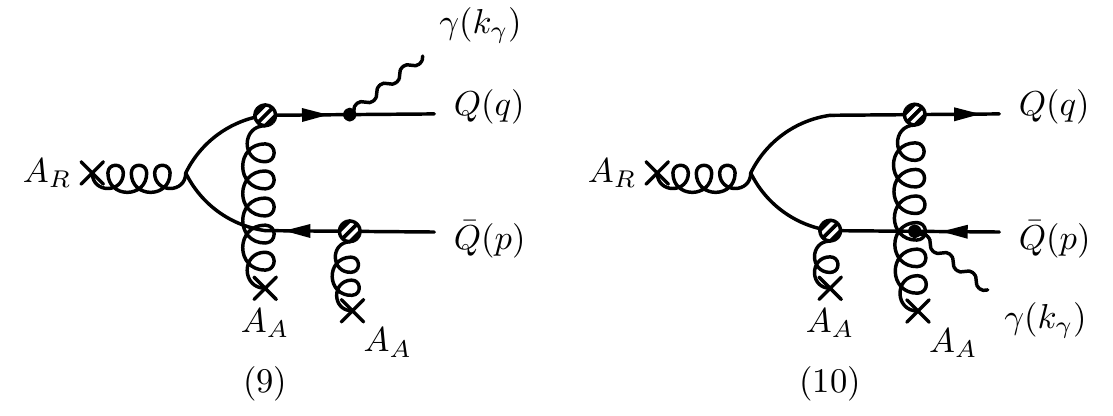}
  \includegraphics[scale=1.0]{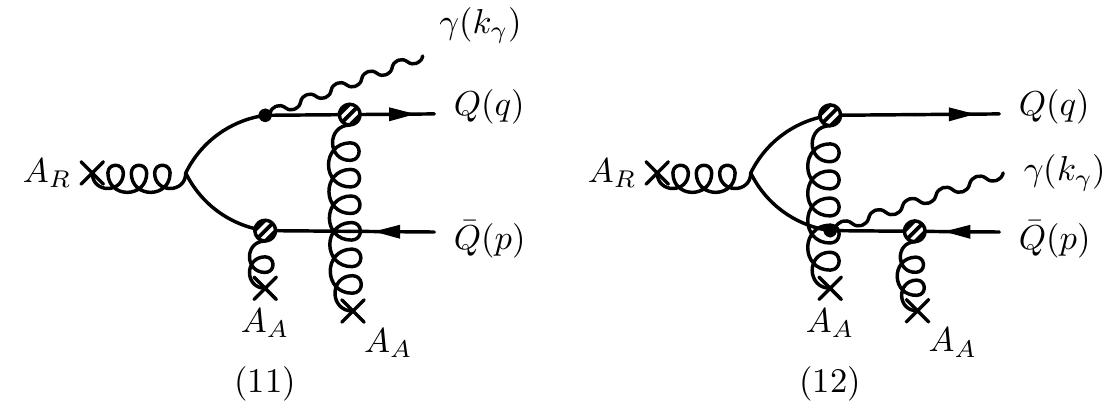}
  \end{center}
  \caption{Regular contributions for the amplitude with two Wilson
    lines in the fundamental representation.}
  \label{fig:D2}
\end{figure}

Let us now consider the case where there are two insertions of the effective vertex on the quark propagator.  The contribution corresponding to a 
photon emission between two insertions of the effective  vanishes for the same
kinematic reasons as previously -- the pole integration yields a null
contribution. Thus the only non-zero contributions come from diagrams where one insertion is on the quark line and the other on the anti-quark line. 
There are four such contributions, which are listed as diagrams
(9)--(12) in Fig.~\ref{fig:D2}.

All of these diagrams are computed with the same logic as previously.
As an example, we focus on diagram (9). The corresponding amplitude can be written as
\begin{equation}
  \begin{split}
  \calM_9^\mu(\MDEP) &= \int\frac{\rmd^4 k}{(2\pi)^4}
    \frac{\rmd^4 k_1}{(2\pi)^4}\ubar(-\rmi q_f e\gamma^\mu) S_0(q+k_\gamma)\calT(k,q+k_\gamma) S_0(q+k_\gamma-k)\\ 
&\times (-\rmi g\slashed{A}_R(k_1)\cdot t) \, S_0(q+k_\gamma-k-k_1)\calT(P - k - k_1,q+k_\gamma-k-k_1)\vp\,.
  \end{split}
\label{eq:m9}
\end{equation}
As for the case with only one effective vertex, and for the same
reasons articulated there, only the proton piece $A_p^\mu$ of the
regular field $A_R^\mu$ contributes.  
The integrals over $k^+$, $k_1^+$ and $k_1^-$ can be performed thanks to the $\delta$-functions in the two effective vertices $\calT$ and in $A_p^\mu$, respectively.
The remaining integration over $k^-$ can be evaluated by the method of residues. 
As in the case of the one
effective vertex insertion, the gluon scattering vertices ($C^\mu_U$ and
$C^\mu_V$) are kinematically forbidden by the pole integration.
Performing similar steps for the remaining diagrams (10)--(12),
the amplitude of the sum of the diagrams (9)--(12) can be expressed as 
\begin{equation}
  \begin{split}
\sum_{\beta=9}^{12}\calM_\beta^\mu(\MDEP)  &= -q_f e g^2\int_{\kp\khp}\int_{\xp\yp}
    \frac{\rhop^a(\khp)}{\khp^2}\,
    \rme^{\rmi\kp\cdot\xp+\rmi(\Pp-\kp-\khp)\cdot\yp} \\
    &\quad\times \ubar T_{q\qbar}^\mu(\kp,\khp)\bigl[\Uf(\xp)-1\bigr]
    t^a\bigl[\Uf^\dag(\yp)-1\bigr]\vp\,,
  \end{split}
\label{eq:MQQ}
\end{equation}
where
\begin{equation}
  T_{q\qbar}^\mu(\kp,\khp) \equiv \sum^{12}_{\beta=9} R_\beta^\mu(\kp,\khp)\,.
\label{eq:Tqqbar}
\end{equation}
The
Dirac structures here 
for $\beta=9,\dots,12$, corresponding to the diagrams
(9)--(12) in Fig.~\ref{fig:D2}, are
\begin{align}
  R_9^\mu(\kp,\khp) &\equiv -\gmu \qkpropD\gp\frac{(\qps+\kgps-\kps+m)\gm
    (\qps+\kgps-\kps-\khps+m)\gp}{N_k(\kp,\khp)}\,,\notag\\
  R_{10}^\mu(\kp,\khp) &\equiv \frac{\gp(\qps-\kps+m)\gm(\qps-\kps-\khps+m)
    }{N_q(\kp,\khp)}\gp \pkpropD\gmu\,,\notag\\
  R_{11}^\mu(\kp,\khp) &\equiv 2p^+\frac{\gp(\slashed{q} + m -\kps)\gmu
    (\qps+\kgps-\kps+m)\gm(\qps+\kgps-\kps-\khps+m)\gp}
    {S(\kp,\khp)N_k(\kp,\khp)} \notag\\
  &\quad -M^2(\qp+\kgp-\kp-\khp) \frac{\gp(\slashed{q}+m-\kps)\gmu\gp\gm
    (\qps+\kgps-\kps-\khps+m)\gp}{S(\kp,\khp)N_k(\kp,\khp)}\,,\notag\\
  R_{12}^\mu(\kp,\khp) &\equiv 2q^+\frac{\gp(\qps-\kps+m)\gm(\qps-\khps-\kps+m)
    \gmu(\Pps-\khps-\kps-\slashed{p}+m)\gp}{S(\kp,\khp)N_q(\kp,\khp)} \notag\\
  &\quad +M^2(\qp-\kp)\frac{\gp(\qps-\kps+m)\gm\gp\gmu
    (\Pps-\khps-\kps-\slashed{p}+m)\gp}{S(\kp,\khp)N_q(\kp,\khp)}\,.
\label{eq:MQ-beta}
\end{align}
For clarity of presentation, the following functions in the
denominator have been defined as
\begin{equation}
  \begin{split}
    N_q(\kp,\khp) &\equiv 2(p^++k_\gamma^+) M^2(\qp-\kp)
      + 2q^+ M^2(\qp-\kp-\khp)\,,\\
    N_k(\kp,\khp) &\equiv 2p^+ M^2(\qp+\kgp-\kp)
      + 2(q^++k_\gamma^+) M^2(\qp+\kgp-\kp-\khp)\,,\\
    S(\kp,\khp) &\equiv 4p^+q^+k_\gamma^- + 2q^+
      M^2(\qp+\kgp-\kp-\khp) + 2p^+ M^2(\qp-\kp)\,.
  \end{split}
\label{eq:MQ-denom}
\end{equation}

In all the equations presented here, $\kp$ stands for the momentum
transferred from the dense nucleus to the quark line.  Likewise,
$\khp$ is the momentum transferred from the proton. This is transparent 
since the proton color sources $\rhop^a(\khp)$ are dependent on this integration
variable alone.  From this fact, and from the fact
that the initial momentum flow must be inferred from the final state
momenta $\Pp$, one can readily notice that $\Pp-\kp-\khp$ is 
the momentum transfer from the nucleus to the antiquark.

\subsection{Singular contributions to the amplitude}
\label{sec:SD}
\begin{figure}
  \begin{center}
  \includegraphics[scale=1.0]{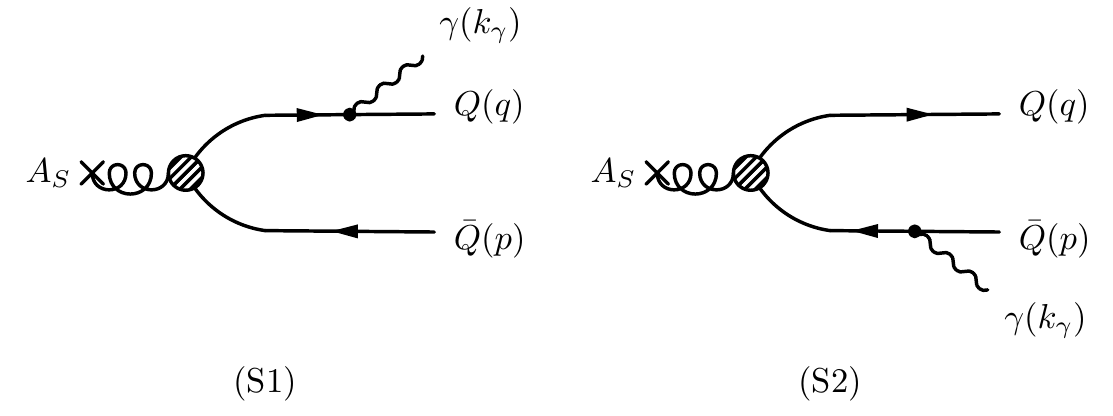}
  \end{center}
  \caption{Singular diagrams of the process.  The blob represents the
    effective vertex defined in Eq.~\eqref{eq:fundvert}.}
  \label{fig:DS}
\end{figure}

The terms for which the $q\qbar$ is produced by the singular part of
the field $A_S^\mu$ are represented by the diagrams (S1) and (S2) in
Fig.~\ref{fig:DS}, 
corresponding to both the
creation of the $q\qbar$ pair, and the subsequent emission of the
photon from within the nucleus.  The amplitude for this process is
non-vanishing in the Lorenz gauge $\partial_\mu A^\mu=0$ and needs further explanation.  Without regularization, the
expressions for amplitudes (S1) and (S2) would look the same as in
Eq.~\eqref{eq:MA1}, but with the regular field exchanged for the
singular field.  With regularization, the $\delta(x^+)$ function has a small
width: $\delta(x^+)\rightarrow \delta_\varepsilon(x^+)$.  This
allows the $q\qbar$ to undergo multiple gluon scatterings.

Diagrams (S1) and (S2) split into four different parts;  each one
corresponds to insertions on the quark and the antiquark lines, making
four distinct combinations.  However these insertions have to be
treated differently than in the previous terms we considered as they
occur inside the regularized region.  This can be achieved by changing
the Wilson lines in the insertions into incomplete Wilson lines.
Summing all the terms that make up (S1), we find,
\begin{equation}
  \calM_{S 1}^\mu(\MDEP) = \int\rmd^4x\, \rme^{\rmi P\cdot x}\,
    \ubar (-\rmi q_f e\gmu) S_0(q + k_\gamma)\Uf(\infty,x^+,\xp)
    (-\rmi g\slashed{A}_S(x)\cdot t )\,\Uf^\dag(\infty,x^+,\xp)\,\vp\,.
\label{eq:sing}
\end{equation}
Several points have to be noted about this expression.  Firstly, the structure of the amplitude reflects the fact that for these diagrams the 
quarks do not rescatter again  as a consequence of the vanishing duration of the interaction in the
$\varepsilon\rightarrow 0$ limit.  In this limit the insertion of the
singular field and the rescattering occur in the same transverse
plane, which is intuitively what one would expect from a $q\qbar$ pair
being created and interacting inside a heavily boosted nucleus.
Secondly, the photon emission from inside the nucleus or followed by
another scattering would be tantamount to resolving the nuclear gluon
shock wave; this too is not kinematically viable.  Therefore one can only
have the addition of an external photon leg emitted by the outgoing
quark or antiquark.

Using the identity
\begin{equation}
  \Uf(\infty,x^+,\xp) t^a\Uf^\dag(\infty,x^+,\xp) = t^b U^{ba}(\infty,x^+,\xp)\,,
\end{equation}
and the formula~\cite{Blaizot:2004wv}
\begin{equation}
  \frac{\rmi g}{2}\int_{-\infty}^{\infty} \rmd x^+ \bigl[
    U(\infty,x^+;\xp)\,A_A^-(x)\cdot T\,V(\infty,x^+;\xp)\bigr]
  = U(\xp) - V(\xp)\,,
\end{equation}
we obtain,
\begin{equation}
  \begin{split}
  \calM_{S 1}^\mu(\MDEP) = -\frac{q_f e g^2}{P^+}\int_{\khp}\int_{\xp}
  \frac{\rhop^a(\khp)}{\khp^2}\, \rme^{\rmi(\Pp-\khp)\cdot\xp} &
  \bigl\{ [U(\xp)-1] - [V(\xp)-1]\bigr\}^{ba} \\
  &\times \ubar\gmu \qkpropD\gp t^b\vp\,.
  \end{split}
\label{eq:MS1}
\end{equation}
A similar procedure for (S2) gives
\begin{equation}
  \begin{split}
  \calM_{S 2}^\mu(\MDEP) = \frac{q_f e g^2}{P^+}\int_{\khp}\int_{\xp}
  \frac{\rhop^a(\khp)}{\khp^2}\, \rme^{\rmi(\Pp-\khp)\cdot\xp} &
  \bigl\{ [U(\xp)-1] - [V(\xp)-1] \bigr\}^{ba} \\
  &\times \ubar\gp \pkpropD\gmu t^b\vp\,.
  \end{split}
\label{eq:MS2}
\end{equation}
In these expressions, we have added and substracted the adjoint
representation identity matrix to show their similarity to
Eqs.~\eqref{eq:MR1} and \eqref{eq:MR2}.  In the following, it will be
shown that some of these contributions cancel out.

\subsection{Assembling the contributions: the complete result for the photon amplitude}

The net contribution to the amplitude is given by a summation of the terms in Eqs.~\eqref{eq:MR1}, \eqref{eq:MR2}, \eqref{eq:qq}, \eqref{eq:MQQ}, \eqref{eq:MS1} and \eqref{eq:MS2}.
There are several cancellations that occur when we put
these terms
together.
The cancellation of
the Wilson line $V(\xp)$ will be addressed first, as it was
anticipated, and it stands as a good check of our  computation.  
Firstly,  using the
relations of Eqs.~\eqref{eq:Lip1} and \eqref{eq:reg}
$C_{V,\text{reg}}^\mu(P)$ is explicitly written as
\begin{equation}
  C_{V,\text{reg}}^\mu(P) = 2P^\mu -\frac{P^2}{P^+}\delta^{\mu-}\,.
\label{eq:CVreg}
\end{equation}
Using the identities
\begin{equation}
  \begin{split}
  & \ubar\gmu \qkpropD\slashed{P} t^a\vp
    = \ubar\gmu t^a\vp\,,\\ 
  & \ubar\slashed{P}\pkpropD\gmu t^a\vp
    = \ubar\gmu t^a\vp\,,
  \end{split}
\label{eq:Canc1}
\end{equation}
it follows that the first term of \eqref{eq:CVreg} cancels out within Eqs.~\eqref{eq:MR1} and \eqref{eq:MR2}.
These relations leave us with only the second term in $C_{V,\text{reg}}^\mu(P)$. 
It is straightforward to check that both remaining
contributions in Eqs.~\eqref{eq:MR1} and \eqref{eq:MR2} are the
counterparts of the $V$ dependent terms in the singular diagrams, and
so Eqs.~\eqref{eq:MS1} and \eqref{eq:MS2} demonstrate the anticipated
cancellation.

The resulting effective vertex for the $U$ terms in the sum of
the diagrams (R1), (S1) and separately (R2), (S2) is
\begin{equation}
  \slashed{C}_U(P,\khp) - \gp \frac{1}{P^+}\,,
\label{eq:eff}
\end{equation}
which can equivalently be written as
\begin{equation}
  \slashed{C}_U(P,\khp) + \frac{1}{2}\slashed{C}_V(P)
  = \Lip(P,\khp)\,,
\end{equation}
where the expression $C_L^\mu$ 
is the Lipatov effective
vertex
\begin{equation}
  C_L^+(q,\khp) = q^+-\frac{\khp^2}{q^-+\rmi\epsilon}\,, \quad
  C_L^-(q,\khp) = \frac{(\qp-\khp)^2}{q^+}-q^-\,, \quad
  \bC_{L\perp}(q,\khp) = \qp - 2\khp\,.
\label{eq:Lip2}
\end{equation}
 
The terms that survive the above mentioned cancellation give the net
amplitude for a gluon to first scatter off the nucleus before emitting
a $q\qbar$ pair and can be expressed as
\begin{equation}
\begin{split}
\calM_\beta^\mu(\MDEP) &\equiv \calM_{R\beta}^\mu(\MDEP) + \calM_{S\beta}^\mu(\MDEP)\\
&= -q_f e g^2\int_{\khp}\int_{\xp}\frac{\rhop^a(\khp)}{\khp^2}\,
\rme^{\rmi(\Pp-\khp)\cdot \xp}\ubar [U(\xp)-1]^{ba} t^b R^\mu_\beta(\khp)\vp\,,
\end{split}
\end{equation}
with $\beta = 1,2$ and where
\begin{equation}
\begin{split}
& R^\mu_1(\khp) \equiv -\gmu \qkpropD\frac{\Lip(P,\khp)}{P^2}\,,\\
&R_2^\mu(\khp) \equiv \frac{\Lip(P,\khp)}{P^2}\pkpropD\gmu\,.
\end{split}
\end{equation}
Combining all the results,
the full amplitude vector is 
\begin{equation}
  \begin{split}
    \calM^\mu(\MDEP) &= \sum_{\beta=1}^{12}\calM_\beta^\mu(\MDEP) = -q_f e g^2\int_{\kp,\khp}\int_{\xp\yp}
    \frac{\rhop^a(\khp)}{\khp^2}\,
    \rme^{\rmi\kp\cdot\xp + \rmi(\Pp-\kp-\khp)\cdot\yp} \\
    &\qquad \times \ubar \Bigl\{ T_g^\mu(\khp)[U(\xp)-1]^{ba} t^b
    + T_q^\mu(\khp)[\Uf(\xp)-1] t^a \\
    &\qquad + T_{\qbar}^\mu(\khp) t^a [\Uf^\dag(\yp)-1]
    + T_{q\qbar}^\mu(\kp,\khp)[\Uf(\xp)-1]t^a[\Uf^\dag(\yp)-1]\Bigr\}\,,
  \end{split}
\label{eq:ampfirst}
\end{equation}
with 
\begin{equation}
  T_g^\mu(\khp) \equiv \sum_{\beta=1}^2 R^\mu_\beta(\khp)\,.
\end{equation}
This last expression extends the Dirac structure
found in Ref.~\cite{Blaizot:2004wv} to photon production.
We note that in Eq.~\eqref{eq:ampfirst} we introduced a dummy integration over $\xp$, $\yp$ and $\kp$ to ensure that all the terms have identical integration variables.

The result \eqref{eq:ampfirst} can be further simplified by making use of the identities,
\begin{equation}
  \begin{split}
  & T_g^\mu(\Pp) + T_q^\mu(\Pp) + T_{\qbar}^\mu(\Pp)
    - T_{q\qbar}^\mu(\mathbf{0},\Pp) = 0\,,\\
  & T_q^\mu(\Pp-\kp)-T_{q\qbar}^\mu(\kp,\Pp-\kp) = 0\,, \qquad
    T_{\qbar}^\mu(\khp) - T_{q\qbar}^\mu(\mathbf{0},\khp) = 0\,.
  \end{split}
\label{eq:amp-ident}
\end{equation}
Using Eq.~\eqref{eq:amp-ident}, the expression for the amplitude can considerably simplify to
\begin{equation}
  \begin{split}
    \calM^\mu(\MDEP) &= -q_f e g^2\int_{\kp \khp}\int_{\xp\yp}
    \frac{\rhop^a(\khp)}{\khp^2}\,
    \rme^{\rmi\kp\cdot\xp+\rmi(\Pp-\kp-\khp)\cdot\yp} \\
    &\qquad\times \ubar \bigl\{ T_g^\mu(\khp)U(\xp)^{ba}t^b
    +T_{q\qbar}^\mu(\kp,\khp)\Uf(\xp)t^a\Uf^\dag(\yp)\bigr\}\vp\,.
  \end{split}
\label{eq:full-amp}
\end{equation}
This expression for the photon amplitude is a key result of this
work. In \ref{app:LC}, we will show that this expression for the
amplitude in Lorenz gauge is identical to the expression derived in
 light-cone gauge.

Before we conclude this section, we wish to make a few points
regarding the final result. Firstly, in Eq.~(\ref{eq:amp-ident}), the
sum of the four effective vertices is zero as a consequence of
momentum conservation. If there is no nuclear and proton momentum
transfer, the quark-antiquark dipole cannot be created.
The second and third relations in Eq.~(\ref{eq:amp-ident}) stand for the vanishing of
  contributions if there is no momentum transfer either from the
  projectile or the target. 
We are thus left only with contributions to the photon amplitude that
have i) both the quark and the antiquark interact with the nucleus
after being created (and before or after radiating the photon), and
those ii) where the gluon from the proton scatters off the nucleus
before creating the quark-antiquark pair and thence, the photon.

\section{The inclusive photon cross section}
\label{sec:CS}

The probability for creating a $q\qbar$ pair with 4-momenta $q$ and
$p$, respectively, and a photon with momentum $k_\gamma$, for a fixed
distribution of sources $\rhop$ in the projectile and $\rhoA$ in the
target, respectively, is given by
\begin{equation}
  P_{\rm incl.}^\gamma[\rhop,\rhoA] = \int
  \frac{\rmd^3\bp}{(2\pi)^3 2E_p} \, \frac{\rmd^3\bq}{(2\pi)^3 2E_q}\,\frac{\rmd^3\kg}{(2\pi)^3 2E_{k_\gamma}}\,
  \sum_\lambda\sum_{\text{spin}}\bigl|\calM_\lambda(\MDEP) \bigr|^2\,.
\end{equation}
Here $E_p$, $E_q$ and $E_{k_\gamma}$ denote the relativistic energies of the antiquark, quark and photon, respectively. 
The sum over polarizations can be taken by noting that
\begin{equation}
  \sum_\lambda \bigl|\calM_\lambda(\MDEP) \bigr|^2 = \sum_\lambda
  \epsilon_\mu(\kg,\lambda)\epsilon_\mu^\ast(\kg,\lambda)
  \calM^\mu(\MDEP)\,\calM^{\mu\ast}(\MDEP)\,.
\label{eq:pos}
\end{equation}
The color average of an inclusive quantity $\calO$ must be taken after
taking the modulus squared of the amplitude~\cite{Blaizot:2004wu},
\begin{equation}
  \langle\calO\rangle = \int\calD\rhop\,\calD\rhoA\,
  W_p[\rhop]\,W_A[\rhoA]\,\calO[\rhop,\rhoA]\,.
\label{eq:col-avg}
\end{equation}
The weight functionals $W_p[\rhop]$ and $W_A[\rhoA]$ are density
matrices that obey the JIMWLK evolution
equations~\cite{JalilianMarian:1997jx, JalilianMarian:1997dw,
  Iancu:2000hn,Iancu:2001ad} that describe the renormalization group
evolution of distributions of color charges in the wave-functions of
the projectile and the target, respectively, from their respective
fragmentation regions at large $x$ down to the small $x$ values probed
by measurements in high energy collisions.

Thus the impact parameter-dependent triple differential probability
corresponding to the probability functional defined above is
\begin{equation}
  \frac{\rmd P^\gamma(\bperp)}{\rmd^6 K_\perp \rmd^3\eta_K}
  = \frac{1}{8(2\pi)^9}\sum_\lambda\sum_{\text{spin}}
  \left\langle \bigl| \calM_\lambda(\MDEP) \bigr|^2 \right\rangle\,.
\label{eq:trip-diff-form}
\end{equation}
where $\rmd^6 K_\perp\equiv \rmd^2\pp\rmd^2\qp\rmd^2\kgp$  and
$\rmd^3\eta_K\equiv \rmd\eta_p \rmd\eta_q\rmd \eta_{k_\gamma}$.  The
angle brackets, $\langle\cdots\rangle$, represent the color average in
Eq.~\eqref{eq:col-avg}.  The triple-differential inclusive
cross section can be obtained by integrating the above expression over
impact parameter $\bperp$,
\begin{equation}
  \TCS = \int_{\bperp} \frac{\rmd P^\gamma(\bperp)}{\rmd^6K_\perp\rmd^3\eta_K}\,.
\end{equation}
Using  Eqs.~\eqref{eq:full-amp} and \eqref{eq:trip-diff-form}, the
triple differential probability can be expressed as
\begin{align}
  \frac{\rmd P^\gamma(\bperp)}{\rmd^6 K_\perp \rmd^3\eta_K} &=
  \frac{q_f^2 e^2 g^4}{8(2\pi)^9} \int_{\kp\kp'\khp\khp'}\int_{\xp\xp'\yp\yp'}
  \rme^{\rmi(\kp\cdot\xp-\kp'\cdot\xp')+\rmi(\Pp-\kp-\khp)\cdot\yp-\rmi(\Pp-\kp'-\khp')\cdot\yp'} \notag\\
  &\quad\times \frac{\left\langle\rhop^a(\khp)\rhop^{\dag a'}(\khp')\right\rangle}
    {\khp^2\khp^{\prime 2}} \Bigl\{ \tau_{g,g}(\khp;\khp')
    \tr_c\bigl\langle t^b U^{ba}(\xp)t^{b'} U^{\dag a'b'}(\xp')\bigr\rangle
    \notag\\
  &\qquad + \tau_{q\qbar,g}(\kp,\khp;\khp')
    \tr_c\bigl\langle \Uf(\xp)t^a\Uf^\dag(\yp)t^{b'} U^{\dag a'b'}(\xp')\bigr\rangle
    \notag\\
  &\qquad + \tau_{g,q\qbar}(\khp;\kp',\khp')
    \tr_c\bigl\langle t^b U^{ba}\Uf(\yp')t^{a'}\Uf^\dag(\xp')\bigr\rangle
    \notag\\
  &\qquad + \tau_{q\qbar,q\qbar}(\kp,\khp;\kp',\khp')
    \tr_c\bigl\langle \Uf(\xp)t^a\Uf^\dag(\yp)\Uf(\yp')t^{a'}\Uf^\dag(\xp')\bigr\rangle\Bigr\}\,.
\label{eq:Full2}
\end{align}
where we wrote four Dirac traces in a compact form using the following
notation,
\begin{equation}
  \tau_{n,m} \equiv \tr\bigl[ (\slashed{q}+m)T_n^\mu(m-\slashed{p})
    \go T'^{\dag}_{m,\mu}\go \bigr]\,.
\end{equation}
with $n,m\in\{g,q\qbar\}$ and where we used the abbreviations $T^\mu_g \equiv T_g^\mu(\khp)$, $T'^{\mu}_g \equiv T_g^{\mu}(\khp')$ and $T^\mu_{q\qbar} \equiv T_{q\qbar}^\mu(\khp,\kp)$, $T'^{\mu}_{q\qbar} \equiv T^\mu_{q\qbar}(\khp',\kp')$.

We will now shift the spatial transverse coordinates by the impact
parameter to make them relative to the center of the nucleus,
\begin{equation}
  \{\xp,\xp',\yp,\yp'\}\; \to\;
  \{\xp-\bperp,\;\xp'-\bperp,\;\yp-\bperp,\;\yp'-\bperp\}\,.
\label{eq:shift}
\end{equation}
Since the Wilson line correlators are approximately translational
invariant for a large nucleus, the phases in the integrand yield a
factor of $\rme^{\rmi\bperp\cdot(\khp-\khp')}$ after the above
coordinate shift. Thus, to obtain the triple-differential
cross section we can easily integrate over the impact parameter
$\bperp$, resulting in a $\delta$-function: $(2\pi)^2\delta^{(2)}(\khp'-\khp)$.
In Eq.~\eqref{eq:Full2}, we also introduce a dummy integration over $\kAp$, together with a $\delta$-function representing overall momentum conservation $(2\pi)^2\delta^{(2)}(\Pp-\khp-\kAp)$.

As discussed previously in
Refs.~\cite{Blaizot:2004wu,Blaizot:2004wv}, the color averages in the
formula above can be re-expressed in terms of novel unintegrated
distribution functions.  The correlator of color sources in the
projectile proton is defined to be
\begin{equation}
  \langle\rhop^{a}(\khp)\rhop^{\dag b}(\khp')\rangle
  = \frac{\delta^{ab}}{2\pi \NC C_F\, g^2}
  \biggl(\frac{\khp+\khp'}{2}\biggr)^2 \int_{\rp}\rme^{\rmi(\khp-\khp')\cdot \rp}\,
  \frac{\rmd\varphi_p\bigl(\tfrac{1}{2}(\khp+\khp')|\rp\bigr)}{\rmd^2\rp}\,,
\end{equation}
where $\rmd\varphi_p/\rmd^2\rp$ is the proton unintegrated gluon
distribution per unit area, $\rp$ is the variable that runs over the
transverse profile of the proton, $C_F\equiv(\NC^2-1)/2\NC$ is the
SU($\NC$) quadratic Casimir in the fundamental representation.  When
the momenta in the argument of $\rmd\varphi_p/\rmd^2\rp$ are
identical as is the case after the integration over $\bperp$, the
expression greatly simplifies to
\begin{equation}
  \langle\rhop^a(\khp)\rhop^{\dag b}(\khp)\rangle
  = \frac{\delta^{ab}\khp^2}{2\pi \NC C_F\, g^2}\,\varphi_p(\khp,Y_p)\,.
\label{eq:rho_corrproton}
\end{equation}
with $\varphi_p(\khp,Y_p)$ being the unintegrated gluon distribution in the proton.
In line with the dilute-dense expansion performed here to $O(\rho_p^1)$, $\varphi_p(\khp)$ will evolve according to the BFKL equation.

The identification of the correlator of color sources can be
straightforwardly generalized to correlators of Wilson lines which
likewise can be expressed in terms of nuclear unintegrated
distribution
functions~\cite{Blaizot:2004wv,Fujii:2005vj,Fujii:2006ab}.  In
conventional pQCD language, these unintegrated momentum
distributions resum a sub-class of all twist correlations in the
nucleus.  The correlator of two adjoint Wilson lines can be expressed
as
\begin{equation}
  \begin{split}
  & \int_{\kp\kp'}\int_{\xp\xp'\yp\yp'}
  \rme^{\rmi(\kp\cdot\xp-\kp'\cdot\xp')+\rmi(\kAp-\kp)\cdot\yp-\rmi(\kAp-\kp')\cdot\yp'}\,
  \delta^{aa'}\,\tr_c\bigl\langle t^b U^{ba}(\xp)t^{b'} U^{\dag a'b'}(\xp')\bigr\rangle\\
  &\equiv \frac{2\NC\alpha_S}{\kAp^2}\,\phi_A^{g,g}(\kAp)\,.
  \end{split}
\label{eq:2Corr}
\end{equation}
Similarly, the three point fundamental-adjoint Wilson line correlator
can be expressed as
\begin{equation}
  \begin{split}
  & \int_{\kp'}\int_{\xp\xp'\yp\yp'}
  \rme^{\rmi(\kp\cdot\xp-\kp'\cdot\xp')+\rmi(\kAp-\kp)\cdot\yp-\rmi(\kAp-\kp')\cdot\yp'}\,
  \delta^{aa'}\,\tr_c\bigl\langle \Uf(\xp)t^a\Uf^\dag(\yp)t^{b'} U^{\dag a'b'}(\xp')\bigr\rangle\\
  &\equiv \frac{2\NC\alpha_S}{\kAp^2}\,\phi_A^{q\qbar,g}(\kp,\kAp-\kp;\kAp)\,,
\label{eq:3Corr}
  \end{split}
\end{equation}
and likewise for its Hermitean conjugate expression in the
cross section.  Finally, the four point correlator of fundamental
Wilson lines can be expressed as
\begin{equation}
  \begin{split}
  & \int_{\xp\xp'\yp\yp'}
  \rme^{\rmi(\kp\cdot\xp-\kp'\cdot\xp')+\rmi(\kAp-\kp)\cdot\yp-\rmi(\kAp-\kp')\cdot\yp'}\,
  \delta^{aa'}\,\tr_c\bigl\langle \Uf(\xp)t^a\Uf^\dag(\yp)\Uf(\yp')t^{a'}\Uf^\dag(\xp')\bigr\rangle\\
  &\equiv \frac{2\NC\alpha_S}{\kAp^2}\,\phi_A^{q\qbar,q\qbar}(\kp,\kAp-\kp;\kp',\kAp-\kp')\,.
  \end{split}
\label{eq:4Corr}
\end{equation}
The correlators can be evaluated for a large nuclei with Gaussian random
color sources~\cite{McLerran:1993ni,McLerran:1993ka,McLerran:1994vd}
that satisfy the relation,
\begin{equation}
  \langle \rhoA^a(\xp)\rhoA^b(\yp) \rangle
  = \delta^{ab}\delta^{(2)}(\xp-\yp)\mu_A^2\,,
\label{eq:rho-corr}
\end{equation}
where the color charge squared per unit area is
$\mu_A^2 = A/(2\pi R^2) \sim A^{1/3}$.
In general, the correlations amongst 
color sources in the nuclear wave-function is given by the weight
functional $W_A[\rhoA]$, which as we noted previously, satisfies the
JIMWLK equation.  This equation is formally equivalent to the
Balitsky-JIMWLK hierarchy~\cite{Balitsky:1995ub,Weigert:2000gi} of
$n$-point Wilson line correlators.  The JIMWLK equation has been
solved numerically~\cite{Rummukainen:2003ns,Dumitru:2011vk}.    
It was
found, to a good approximation, that the solution is well represented
by a non-local Gaussian~\cite{Dumitru:2011vk}, with
$\mu_A^2\rightarrow \mu_A^2(Y_A,\xp)$, with which we can define,
\begin{equation}
  \phi_A(Y_A,\kp) = \frac{2\pi \NC C_F \, g^2}{\kp^2}
  \int_{\xp} \mu_A^2(Y_A,\xp) \,,
\label{eq:nonlocalMV}
\end{equation} 
such that $\phi_A(Y_A,\kp)$ obeys the Balitsky-Kovchegov (BK)
equation~\cite{Balitsky:1995ub,Kovchegov:1999yj}.  In this expression,
$Y_A$ is the rapidity, relative to the nuclear beam rapidity, of
gluons from the target off which the $q\qbar$ pair scatters.  The BK
equation, for large $\NC$, is the closed form equation for the
``dipole'' correlator of Wilson lines -- the lowest term in the
Balitsky-JIMWLK hierarchy.  In the low density limit the BK equation goes to the BFKL equation.  Even though the BK equation has a closed
form, it cannot be solved analytically.  It can however be solved
numerically; we will return to this discussion when we later outline
the necessary ingredients for the numerical solution of
Eq.~\eqref{eq:Full4}.

Returning to the triple-differential cross section, with the help of the relations Eq.~\eqref{eq:shift}-\eqref{eq:4Corr}, we express its final form as
\begin{equation}
  \begin{split}
  \TCS &= \frac{\alpha_e\alpha_S^2\,q_f^2}{256\pi^8 C_F}
  \int_{\khp\kAp} (2\pi)^2\delta^{(2)}(\Pp-\khp-\kAp)\,
  \frac{\varphi_p(\khp)}{\khp^2\kAp^2} \\
  &\times \biggl\{
  \tau_{g,g}(\khp;\khp)\phi_A^{g,g}(\kAp)
  + \int_{\kp} 2\text{Re} \bigl[
    \tau_{g,q\qbar}(\khp;\kp,\khp)\bigr]\,
    \phi_A^{q\qbar,g}(\kp,\kAp-\kp;\kAp) \\
  &+ \int_{\kp\kp'}\tau_{q\qbar ,q\qbar}(\kp,\khp;\kp',\khp)\,
    \phi_A^{q\qbar,q\qbar}(\kp,\kAp-\kp;\kp',\kAp-\kp')\biggr\}\,.
  \end{split}
\label{FullTDCS}
\end{equation}
We remind the reader that $\Pp=\pp+\qp+\kgp$.  
It is worth noting that the r.~h.~s. dependence on 
the external momenta lies only in the overall
$\delta$-function and in the Dirac traces.

The expression in Eq.~\eqref{FullTDCS} would be useful if it were
feasible to measure direct photons in coincidence with a quarkonium
state, for instance the $J/\Psi$ meson.  Alternately, by integrating
over either the quark (antiquark), this cross section would provide
the rate for direct photons measured in coincidence with open charm
(anticharm) states.  These measurements are challenging even at LHC energies.  On the other hand, inclusive prompt photon
differential cross section has already been measured at RHIC in
deuteron-gold collisions~\cite{Russcher:2008zz,Adare:2012vn} and photon measurements can be anticipated at both RHIC and LHC in p+A 
collisions in the near future. 

The simplest quantity that can be measured is the inclusive
prompt-photon single-differential cross section.  For this one must
integrate over the quark and the antiquark momenta and rapidities, and
one obtains,
\begin{align}
  \frac{\rmd\sigma^\gamma}{\rmd^2\kgp \rmd\eta_{k_\gamma}} &=
  \frac{\alpha_e \alpha_S^2 q_f^2}{16\pi^4 C_F}\int_0^\infty\frac{\rmd q^+}{q^+}\,
  \frac{\rmd p^+}{p^+} \int_{\khp\kAp\qp\pp}
  (2\pi)^2\delta^{(2)}(\Pp-\khp-\kAp)\,\frac{\varphi_p(\khp)}{\khp^2\kAp^2}\notag\\
  &\times\biggl\{ \tau_{g,g}(\khp;\khp)\phi_A^{g,g}(\kAp)
 + \int_{\kp} 2\text{Re}\bigl[ \tau_{g,q\qbar}(\khp;\kp,\khp)\bigr]\,
  \phi_A^{q\qbar,g}(\kp,\kAp-\kp;\kAp) \notag\\
  &+ \int_{\kp\kp'} \tau_{q\qbar,q\qbar}(\kp,\khp;\kp',\khp)\,
  \phi_A^{q\qbar,q\qbar}(\kp,\kAp-\kp;\kp',\kAp-\kp')\biggr\}\,.
\label{eq:Full4}
\end{align}
This expression is the main result of this paper.  In the previous
related work on quark-antiquark
production~\cite{Blaizot:2004wv,Fujii:2006ab}, the integration over
the momentum of the quark or the antiquark enabled one to simplify the
result so that it depended only on the $\phi_A^{g,g}$ and
$\phi_A^{q\qbar,g}$ nuclear distributions.  Unfortunately, it appears
no such simplification is possible here due to the particular momentum
dependence of the $\tau_{n,m}$ functions.  Equation~\eqref{eq:Full4},
however, simplifies considerably in the $\kp$ factorization and
collinear pQCD limits, which will be the subject of the next section.

\section{$k_\perp$ factorization and collinear factorization limits}
\label{sec:KF}

Our master expression, Eq.~\eqref{eq:Full4}, can be simplified
considerably in a high transverse momentum expansion of the nuclear
unintegrated distributions.  The high momentum expansion corresponds to expanding the Wilson lines,
$\Uf(\xp)$ and $U(\xp)$ in the fundamental and the adjoint
representations, to lowest non-trivial order in terms of
$\rhoA/\nabla_\perp^2$.  This is equivalent to the leading twist
expansion in pQCD as will become manifest when we consider the
collinear factorization limit of our expressions.  Physically, this
limit corresponds to the dynamics when the density of color sources is large
enough to be represented as classical color charges, but only one of the color charges in this classical
color distribution is resolved by a high transverse momentum probe. Keeping only the leading-twist terms in the expansion of the Wilson
lines in Eq.~\eqref{eq:4Corr},
one can straightforwardly
show that~\cite{Blaizot:2004wv,Fujii:2006ab}
\begin{equation}
  \begin{split}
  & \phi_A^{q\qbar,q\qbar}(Y_A,\kp,\kAp-\kp;\kp',\kAp-\kp')
    = (2\pi)^4 \varphi_A(Y_A,\kAp)
    \biggl\{ \frac{C_F}{\NC} \Bigl[\delta^{(2)}(\kp)\delta^{(2)}(\kp') \\
  &\qquad + \delta^{(2)}(\kAp-\kp)\delta^{(2)}(\kAp-\kp')
    \Bigr] + \frac{1}{2\NC^2} \Bigl[
    \delta^{(2)}(\kp)\delta^{(2)}(\kAp-\kp')
    +\delta^{(2)}(\kAp-\kp)\delta^{(2)}(\kp')\Bigr] \biggr\}\,.
  \end{split}
\end{equation}
Similarly,
\begin{equation}
  \phi_A^{q\qbar,g}(Y_A,\kp,\kAp-\kp;\kAp)
  = \frac{1}{2}(2\pi)^2\varphi_A(Y_A,\kAp)\bigl[
    \delta^{(2)}(\kp) + \delta^{(2)}(\kAp-\kp)\bigr]\,,
\end{equation}
and
\begin{equation}
  \phi_A^{g,g}(Y_A,\kAp) = \varphi_A(Y_A,\kAp)\,,
\end{equation}
In complete analogy with \eqref{eq:rho_corrproton} we have defined
\begin{equation}
  \langle\rhoA^a(\kAp)\rhoA^{\dag b}(\kAp)\rangle
  \equiv \frac{\delta^{ab}\kAp^2}{2\pi \NC C_F\, g^2}\,\varphi_A(Y_A,\kAp)\,,
\label{eq:rho_corrnucleus}
\end{equation}
where the $Y_A$ dependence is now shown explicitly. The unintegrated gluon distribution in the nuclei $\varphi_A(Y_A,\kAp)$ evolves at this level of approximations according to the BFKL equation.

Substituting these leading twist distributions into
Eq.~\eqref{eq:Full4}, we find that the cross section simplifies
greatly and can be expressed as
\begin{equation}
  \TCS = \frac{\alpha_e \alpha_S^2 q_f^2 }{256\pi^8 \NC (\NC^2-1)}\int_{\khp\kAp}
  (2\pi)^2\delta^{(2)}(\Pp-\khp-\kAp)\,
  \frac{\varphi_p(Y_p,\khp)\varphi_A(Y_A,\kAp)}{\khp^2\kAp^2}\,
  \Theta(\khp,\kAp)
\label{eq:FullLT2}
\end{equation}
with
\begin{align}
  & \Theta(\khp,\kAp) \notag\\
  &\equiv \NC^2 \tr \Bigl\{ (\slashed{q}+m)
    \bigl[ T_g^\mu(\khp)+T_q^\mu(\Pp-\kAp) \bigr]
    (m-\slashed{p})\go \bigl[ T_{g,\mu}(\khp)+T_{q,\mu}(\Pp-\kAp)
    \bigr]^\dag\go \Bigr\} \notag\\
  &\quad + \NC^2 \tr \Bigl\{ (\slashed{q}+m)\bigl[
    T_g^\mu(\khp)+T_{\qbar}^\mu(\khp)\bigr]
    (m-\slashed{p})\go \bigl[
    T_{g,\mu}(\khp)+T_{\qbar,\mu}(\khp)
    \bigr]^\dag\go \Bigr\} \notag\\
  &\quad - \tr \Bigl\{ (\slashed{q}+m)\bigl[ T_q^\mu(\Pp-\kAp)
    -T_{\qbar}^\mu(\khp)\bigr](m-\slashed{p})\go
    \bigl[ T_{q,\mu}(\Pp-\kAp)-T_{\qbar,\mu}(\khp) \bigr]^\dag\go\Bigr\}\,.
\label{eq:theta}
\end{align}
The expression in Eq.~\eqref{eq:FullLT2} is the $\kp$ factorized
expression for inclusive photon production in high energy QCD and is
analogous to similar expressions derived previously for gluon
production~\cite{Kharzeev:2003wz,Blaizot:2004wu,Dumitru:2001ux} and
$q\qbar$ pair production~\cite{Blaizot:2004wv} in this framework.
To derive Eq.~\eqref{eq:theta} we have taken advantage of the second line in the Eq.~\eqref{eq:amp-ident} to express $T^\mu_{q\qbar}$ in terms of $T^\mu_q$ or $T^\mu_{\bar{q}}$.

\begin{figure}
  \begin{center}
  \includegraphics[width=\textwidth]{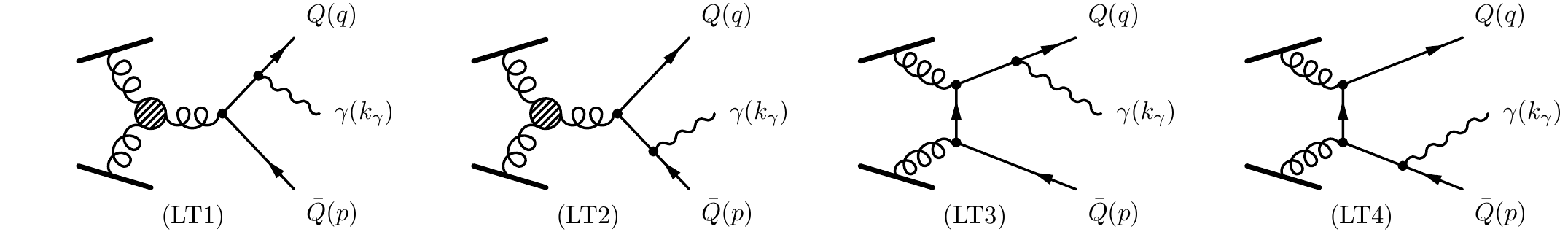}\vspace{1em}
  \includegraphics[width=\textwidth]{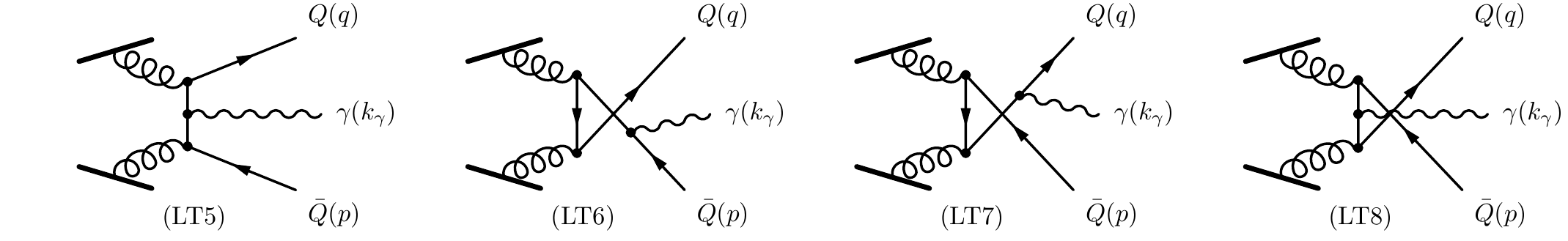}
  \end{center}
  \caption{Leading twist reduction and corresponding diagrams. The upper and lower bold lines correspond to the nuclear and proton sources, respectively. In diagrams LT1 and LT2, the blob represents the Lipatov effective vertex.}
\label{fig:LTDiagrams}
\end{figure}

Alternately, it is useful to arrive at the results in Eqs.~\eqref{eq:FullLT2} and \eqref{eq:theta} starting directly from the perturbative computation of the amplitude.
The leading twist diagrams are listed in Fig.~\ref{fig:LTDiagrams}.
In fact, the leading twist amplitude in Lorenz gauge can immediately be read off from the original amplitudes (1)--(8) by expanding the Wilson lines to the first non-trivial order. In this case, it gives a non-zero contribution to the amplitude at the order $O(\rho_p^1 \rho_A^1)$. The amplitudes (9)--(12) contain two insertions of the effective vertex (see  Eq.~\eqref{eq:fundvert}), and so they do not contribute at the order $O(\rho_p^1 \rho_A^1)$.
We can then write the leading twist amplitude as
\begin{equation}
\calM_{\rm LT} ^\mu(\MDEP) = \int \frac{d^4 k_1}{(2\pi)^4}\int \frac{d^4 k_2}{(2\pi)^4} (2\pi)^4 \delta^{(4)}(P-k_1-k_2) A_{p,\nu_1}^a(k_1)A^b_{A,\nu_2}(k_2) m^{\mu\nu_1\nu_2}_{ab}(k_1,k_2,\MDEP)\,,
\end{equation}
where the gluon fields are given by Fourier transforms of the expressions in Eqs.~\eqref{eq:pA} and \eqref{eq:AA} and
\begin{equation}
m^{\mu\nu_1\nu_2}_{ab}(k_1,k_2,\MDEP) = \sum_{\beta = 1}^8 m^{\mu\nu_1\nu_2}_{\beta,ab}(k_1,k_2,\MDEP)\,,
\end{equation}
with the index $\beta = 1,\dots,8$ corresponding to the respective diagram in Fig.~\ref{fig:LTDiagrams} as
\begin{equation}
\begin{split}
& m^{\mu - + }_{1,ab}(k_1,k_2,\MDEP) = q_f eg^2\ubar \gamma^\mu S_0(q+k_\gamma) \, [t_b,t_a] \, \frac{\slashed{C}_L(P,k_1)}{P^2}\vp\,,\\
& m^{\mu - + }_{2,ab}(k_1,k_2,\MDEP) = q_f eg^2\ubar \, [t_b,t_a]\, \frac{\slashed{C}_L(P,k_1)}{P^2} S_0(-p-k_\gamma) \gamma^\mu\vp\,,\\
& m^{\mu - + }_{3,ab}(k_1,k_2,\MDEP) = -\rmi q_f eg^2\ubar \gamma^\mu S_0(q+k_\gamma) \gamma^+ t_b S_0(k_1 - p) \gamma^- t_a \vp\,,\\
& m^{\mu - + }_{4,ab}(k_1,k_2,\MDEP) = -\rmi q_f eg^2\ubar \gamma^+ t_b S_0(q-k_2) \gamma^\mu S_0(k_1-p) \gamma^- t_a \vp\,,\\
& m^{\mu - + }_{5,ab}(k_1,k_2,\MDEP) = -\rmi q_f eg^2\ubar \gamma^+ t_b S_0(q-k_2) \gamma^- S_0(-p-k_\gamma) \gamma^\mu t_a \vp\,,\\
& m^{\mu - + }_{6,ab}(k_1,k_2,\MDEP) = -\rmi q_f eg^2\ubar \gamma^- t_a S_0(q-k_1) \gamma^+ t_b S_0(-p-k_\gamma) \gamma^\mu \vp\,,\\
& m^{\mu - + }_{7,ab}(k_1,k_2,\MDEP) = -\rmi q_f eg^2\ubar \gamma^- t_a S_0(q-k_1) \gamma^\mu S_0(k_2-p) \gamma^+ t_b \vp\,,\\
& m^{\mu - + }_{8,ab}(k_1,k_2,\MDEP) = -\rmi q_f eg^2\ubar \gamma^\mu S_0(q+k_\gamma) \gamma^- t_a S_0(k_2-p) \gamma^+ t_b \vp\,.
\end{split}
\label{eq:MLTbeta}
\end{equation}
Inserting the gluon fields from Eqs.~\eqref{eq:pA}  and \eqref{eq:AA}, we can perform the light-cone integrations to obtain
\begin{equation}
  \begin{split}
  \calM_{\rm LT}^\mu &= -\rmi q_f e g^2 \int_{\khp\kAp}
  (2\pi)^2\delta^{(2)}(\Pp-\khp-\kAp)\,\frac{\rhop^a(\khp)}{\khp^2}
  \frac{\rhoA^b(\kAp)}{\kAp^2} \\
  &\quad\times \ubar \Bigl[T_g^\mu(\khp)[t_b,t_a]
    + T_q^\mu(\khp)t_b t_a - T_{\qbar}^\mu(\khp)t_a t_b
    \Bigr]\vp\,,
  \end{split}
\label{eq:MLT}
\end{equation}
By taking the modulus squared and by performing the color averages as in Eqs.~\eqref{eq:rho_corrproton} 
and \eqref{eq:rho_corrnucleus}, we can  confirm that
Eq.~\eqref{eq:FullLT2} is reproduced.

The expression for the amplitude in Eq.~\eqref{eq:MLT}, and in particular the individual contributions \eqref{eq:MLTbeta}, can be compared to the amplitude for prompt photon hadroproduction first computed by Baranov et al. \cite{Baranov:2007np} and to the
$Z^0$ hadroproduction from gluon fusion derived recently by Motyka et al. \cite{Motyka:2016lta}.
We have checked that our leading twist $k_\perp$ factorized result agrees exactly with Eq.~(28) of Motyka et al.

One can also take the collinear limit, which as noted previously~\cite{Gelis:2003vh}, can be obtained by taking $\khp\rightarrow 0$ and $\kAp\rightarrow 0$  in the trace element, $\Theta(\khp,\kAp)/\khp^2\kAp^2$, but not in the unintegrated functions. 
This means that the total external momenta vanishes, $\pp+\qp+\kphp = 0$, thus guaranteeing momentum conservation.
The limit
\begin{equation}
\lim_{\substack{\khp \to 0 \\ \kAp \to 0}}\frac{\Theta(\khp,\kAp)}{\khp^2\kAp^2}\,,
\end{equation}
is well defined thanks to the Ward identities
\begin{equation}
k_{1\nu_1}m_{ab}^{\mu\nu_1 +}(k_1,k_2,\MDEP) = 0\, , \qquad k_{2 \nu_2}m_{ab}^{\mu - \nu_2}(k_1,k_2,\MDEP) = 0~,
\end{equation}
ensuring that the amplitude vanishes linearly with $\khp$ and $\kAp$.

The integration over $\khp(\kAp)$ has to be taken then only over the unintegrated distribution functions,
\begin{equation}
x_i f_{g,i}(x_i,Q^2) \equiv \frac{1}{4\pi^3}\int_{0}^{Q^2} \rmd \kp^2 \, \varphi_i(Y_i,\kp) = \frac{1}{\pi^2} \int_{\kp}\varphi_i(Y_i,\kp)\,, \qquad i=p,A\,,
\end{equation}
to get the inclusive photon cross section at fixed $Q^2$ and momentum fractions in the proton  $x_p \sim \frac{Q_S^p}{\sqrt{s}}e^{\eta_{k_\gamma}}$ and in the nuclei $x_A \sim \frac{Q_S^A}{\sqrt{s}} e^{-\eta_{k_\gamma}}$, as 
\begin{equation}
\frac{d\sigma^\gamma}{d^2\kg d\eta_{k_\gamma}} =\frac{1}{16}\int_0^\infty \frac{\rmd q^+}{q^+}\frac{\rmd p^+}{p^+}\int_{\qp \, \pp}(2\pi)^2\delta^{(2)}(\pp+\qp+\kphp) \, x_p f_{g,p}(x_p,Q^2) \, x_A f_{g,A}(x_A,Q^2) \,
 |\mathcal{M}_{gg\to q\qbar\gamma}|^2\,,
 \label{eq:FullColl}
\end{equation}
where the $|\mathcal{M}_{gg\to q\qbar\gamma}|^2$ object is defined as
\begin{equation}
|\mathcal{M}_{gg\to q\qbar\gamma}|^2\equiv\lim_{\substack{\khp \to 0 \\ \kAp \to 0}} \frac{q_f^2 \, \alpha_e\, \alpha_S^2}{N_c(N_c^2 - 1)} \frac{\Theta(\khp,\kAp)}{\khp^2\kAp^2}\,.
\end{equation}
An especially interesting point to note here is the sensitivity of the collinear result in Eq.~\eqref{eq:FullColl} to the nuclear gluon distribution function $f_{g,A}$.
The CGC formalism used in this paper naturally extends beyond this result to the multi-parton distributions of the saturated nucleus as the momentum scales probed approach the semi-hard saturation scale $Q_S^A$.

\section{Summary and Outlook}
\label{sec:SO}

We have computed in this work the leading contribution to inclusive cross section for photon production at central rapidities in high energy proton-nucleus collisions. Our result for the triple differential cross section for a photon accompanied by a quark-antiquark pair is given in Eq.~(\ref{FullTDCS}) and that for the inclusive photon cross section is given in Eq.~(\ref{eq:Full4}). The result in this paper for the class III diagrams, in combination with the result for the class II diagrams in Fig.~\ref{fig:classes} obtained previously in \cite{Benic:2016yqt}, completes the NLO computation of inclusive photon production within the CGC framework. 

The result in Eq.~(\ref{eq:Full4}) has an identical structure to the expression for quark-antiquark pair production computed previously in \cite{Blaizot:2004wv}. Moreover as we show explicitly in Eq.~(\ref{eq:LBK}) of Appendix B, the Low-Burnett-Kroll theorem tells us that, in the soft photon limit, the amplitude for photon production is precisely the amplitude for quark-antiquark pair production multiplied by a simple kinematic expression with the Lorentz structure of the photon. 

The properties of the heavy quark pair production cross section derived in \cite{Blaizot:2004wv} have been explored extensively~\cite{Fujii:2005vj,Fujii:2006ab}. These results, when combined with a CGC+non-relativistic QCD (NRQCD) formalism~\cite{Kang:2013hta}, provide an excellent description of p+p $J/\Psi$ 
production data~\cite{Ma:2014mri} and p+A $J/\Psi$ production data~\cite{Ma:2015sia} at both RHIC and the LHC. The same formalism has also been employed to study $J/\Psi$ production in p+A collisions within a color evaporation model of the hadronization of charm-anticharm pairs to $J/\Psi$ mesons~\cite{Fujii:2013gxa,Ducloue:2015gfa}. 

Computational techniques identical to those employed in these works can be used for quantitative estimates of inclusive photon production in p+A collisions at RHIC and LHC. The momentum dependent multi-parton correlations functions $\phi_A^{g,g}$, $\phi_A^{q\qbar,g}$ and $\phi_A^{q\qbar,q\qbar}$ appearing in Eq.~(\ref{eq:Full4}) can be computed within the non-local Gaussian approximation we alluded to previously; this approximation reproduces a Langevin numerical implementation of the JIMWLK hierarchy~\cite{Dumitru:2011vk} that delivers closed form results for such multi-parton correlators. The large $\NC$ limit of the multi-parton correlators provides an added simplification. Thus while the number of integrals to be performed are greater in the photon case relative to that of $q\qbar$ production, the computation of the former is feasible and will be reported on in future. 

The numerical  results will be very relevant for comparisons to anticipated results for direct photon measurements in p+A collisions at RHIC and the LHC.
As noted, such measurements will be very sensitive to the nuclear gluon distribution function and will provide a quantitative estimate of power corrections to the same. Currently, the deuteron-gold measurements \cite{Russcher:2008zz,Adare:2012vn} at RHIC have covered the kinematic range $ 1 \, {\rm GeV} \lesssim  k_{\gamma\perp} \lesssim 6 \, {\rm GeV}$ with nearly real virtual photons and $ 5 \, {\rm GeV} \lesssim  k_{\gamma\perp} \lesssim 16 \, {\rm GeV}$ with real photons.
It will be important to extend the latter measurements to lower $k_\perp$ to fully explore the gluon saturation regime. 
While the data is in agreement with pQCD, the remaining uncertainty allows for the possibility of thermal photons \cite{Shen:2015qba} on top of the pQCD results. Computations in our framework will help determine whether the latter are necessary. 

Further,  prompt photon measurements will provide an important test of the gluon saturation in general and, in particular, of the multi-parton correlators we have discussed here. They should corroborate the above mentioned studies of quarkonium production and may potentially be more robust since they are less sensitive to hadronization effects. We also note that our framework may provide insight into a long standing experimental puzzle~\cite{Belogianni:2002ic} regarding the ``anomalously" large photon production at soft momenta relative to predictions based on the Low-Burnett-Kroll theorem. We will investigate these issues in future work. 

Another interesting feature of the results presented here is the comparative study of photon production in Lorentz gauge and light-cone gauge. While 
this study is a useful check of our results, it may also have further value in higher order computations that are feasible in this framework.

\section*{Acknowledgements}
S.~B.\ was supported by the European Union Seventh Framework Programme
(FP7 2007-2013) under grant agreement No.\ 291823, Marie Curie
FP7-PEOPLE-2011-COFUND NEWFELPRO Grant No.\ 48.
K.~F.\ was supported by MEXT-KAKENHI Grant No.\ 15H03652 and
15K13479.
R.~V.\ is supported under DOE Contract No. DE-SC0012704. 
This material is also based upon work supported by the U.S. Department of Energy, Office of Science, Office of Nuclear Physics, within the framework of the TMD Topical Collaboration. S.~B. also acknowledges the support of HZZO Grant No. 8799. This work is part of and supported by the DFG Collaborative Research Centre ``SFB 1225 (ISOQUANT)''. K.~F.\ and R.~V.\ would like to thank the Institut f\"{u}r Theoretische Physik at Heidelberg University for kind hospitality during the course of this work and R.~V.\ would like to thank the Excellence Initiative of Heidelberg University for their support. K.~F.\ thanks the Extreme Matter Institute (EMMI) for support. Part of this work comprised the Master's thesis of Garcia-Montero at Heidelberg University; he would like to thank J\"{u}rgen Berges for valuable discusssions and for co-supervising his thesis with R.V. We would also like to thank Werner Vogelsang for useful discussions.

\appendix

\section{Photon amplitude in light-cone gauge}
\label{app:LC}

\subsection{Derivation of the amplitude}

The treatment of the quantum fluctuations would be easier in the
light-cone gauge, $A^+=0$, while the classical solution of the
Yang-Mills equation would be more complicated in the light-cone gauge
having transverse components than in the Lorenz gauge.  Interestingly,
the non-zero component of the classical solution obtained in the
Lorenz gauge is $A^-$ only and $A^+=0$ is consistent with the
light-cone gauge.  Here, we adopt such a hybrid choice of gauge
fixing with the background field in the Lorenz gauge and the quantum
fluctuations in the light-cone
gauge~\cite{Gelis:2005pt,Fukushima:2008ya}.  Then, unlike the Lorenz
gauge, the gauge field from the proton color source does not depend on
$V(\xp)$ but only $U(\xp)$.  We can write it as
\begin{equation}
  A_{\rm LC}^\mu(q) =
    A_{p,\text{LC}}^\mu(q)
    + \frac{\rmi g}{q^2+\rmi q^+\epsilon}
      \int_{\khp}\int_{\xp} \rme^{\rmi (\qp-\khp)\cdot\xp}\,
      C^\mu(q,\khp)\, U(\xp)\, \frac{\rhop(\khp)}{\khp^2}
\label{eq:A1l}
\end{equation}
where the first term represents,
\begin{equation}
  A_{p,\text{LC}}^\pm(q) = 0\,,\qquad
  \boldsymbol{A}_{p,\text{LC}\perp}(q)
  = \frac{\rmi g\, \qp}{(q^++\rmi\epsilon)(q^--\rmi\epsilon)}\,
    \frac{\rhop(\qp)}{\qp^2}\,.
\label{eq:ApLC}
\end{equation}
We note that $A_{p,\text{LC}}$ is non-zero only in the $x^+<0$ region
thanks to the pole at $q^-=\rmi\epsilon$, while the second term in
Eq.~\eqref{eq:A1l} is non-zero only in the $x^+>0$ region.  The
components of $C^\mu(q,\khp)$ in the light-cone gauge are
\begin{equation}
  C^+(q,\khp) = 0\,,\quad
  C^-(q,\khp)
    = \frac{-2\khp\cdot(\qp-\khp)}{q^+ +\rmi\epsilon}\,,\quad
  \bC_\perp(q;\khp)
    = \frac{\qp \, \khp^2}{(q^+ +\rmi\epsilon)(q^- +\rmi\epsilon)} - 2\khp\,.
\end{equation}
Because no $V(\xp)$ appears, we do not have to consider the singular
diagrams.

We can give the counterparts of the amplitude vectors corresponding to the diagrams (R1)--(R2) in
Fig.~\ref{fig:D0} as
\begin{align}
  \calM^\mu_1(\MDEP) &= \ubar (-\rmi q_f e\gmu)
    S_0(q+k_\gamma)(-\rmi g\slashed{A}_{\rm LC}(P)\cdot t)\,\vp\,,\\
  \calM^\mu_2(\MDEP) &= \ubar (-\rmi g\slashed{A}_{\rm LC}(P)\cdot t)\,
    S_0(-p-k_\gamma)(-\rmi q_f e\gmu) \vp\,,
\end{align}
This expression seems to be parametrically identical to that in the
Lorenz gauge but a difference arises from the gauge field from the
proton;  in the Lorenz gauge $A_R^\mu$ appear with $C_U^\mu$ and
$C_V^\mu$ or $C_L^\mu$ after cancellation of $C_V^\mu$, while in the
light-cone gauge $A_{\rm LC}^\mu$ appear with $C^\mu$ as
defined above.
After some calculations
we find,
\begin{equation}
\begin{split}
  \calM_\beta^\mu(\MDEP) &= -q_f e g^2 \int_{\khp}\int_{\xp}
  \frac{\rhop^a(\khp)}{\khp^2}\, \rme^{\rmi(\Pp-\khp)\cdot\xp}
  [U(\xp)-1]^{ba} \ubar R_\beta^{\mu\nu} C_\nu(P,\khp) t^b \vp\,,
\end{split}
\label{eq:m7lc}
\end{equation}
where $\beta=1,2$. 
To facilitate a comparison between different gauges we introduced extended tensors for Dirac indices as  
\begin{equation}
  R_1^{\mu\nu} \equiv -\frac{1}{P^2}\gmu \qkpropD\gnu\,, \qquad  R_2^{\mu\nu} \equiv \frac{1}{P^2}\gnu \pkpropD\gmu\,,
\end{equation}
and for later convenience we also define
\begin{equation}
R^{\mu\nu}_g \equiv \sum_{\beta=1}^2 R_\beta^{\mu\nu}~.
\end{equation}
Now we understand that we can express $T_g^\mu(\khp)$ in the Lorenz gauge as
\begin{equation}
  T_g^\mu(\khp) = R_g^{\mu\nu} C_{L \, \nu}(P,\khp)\,.
\label{eq:tglc}
\end{equation}

Similarly, we proceed to evaluate the counterparts of regular diagrams
(3)--(8) in Fig.~\ref{fig:Reg1}.  For example, let us look carefully at
the evaluation of $\calM_3^\mu$.  The diagram (3) immediately leads to
\begin{equation}
  \calM_3^\mu(\MDEP) = \int\frac{\rmd^4 k_2}{(2\pi)^4}
    \ubar(-\rmi q_f e\gamma^\mu) S_0(q+k_\gamma)
    \calT(q+k_\gamma,k_2) S_0(q+k_\gamma-k_2)(-\rmi g
    \slashed{A}_{\rm LC}(P-k_2)\cdot t)\,\vp\,.
\end{equation}
in which only $A_{p,{\rm LC}}^\mu$ contributes.  
In the above
expression the $k_2^+$ integration is trivial once we insert the effective vertex $\calT(q+k_\gamma,k_2)$, given in \eqref{eq:fundvert}.
The $k_2^-$ integration picks up a singularity in $S_0(q+k_\gamma-k_2)$.
After the variable change change $k_2 = P-k_1$ we find
\begin{equation}
  \calM_3^\mu(\MDEP) = q_f e g^2 \int_{\khp}\int_{\xp}
  \frac{\rhop^a(\khp)}{\khp^2}\, \rme^{\rmi(\Pp-\khp)\cdot\xp}
  \ubar R_3^{\mu i}(\khp) \frac{k_{1i}}{P^+}
  [\Uf(\xp)-1] t^a \vp\,,
\end{equation}
where
\begin{equation}
  R_3^{\mu\nu}(\khp) \equiv \gmu \qkpropD \gp
  \frac{\slashed{k}_1-\slashed{p}+m}{2(q^+ +k_\gamma^+)p^- + M^2(\khp-\pp)}
  \gnu\,.
\end{equation}
We note that in the Lorenz gauge $R_3^{\mu i}(\khp)(k_{1i}/P^+)$ would be
replaced with $-R_3^{\mu -}$.  For diagrams (4)--(8) we can carry out
similar procedures to define $R_\beta^{\mu\nu}(\khp)$ by generalizing $\gm$
in $R_\beta^\mu(\khp)$ to $\gnu$, and then $\calM_\beta^\mu$ is simply
given by $R_\beta^{\mu i}(\khp)(k_{1i}/P^+)$ instead of $-R_\beta^{\mu -}(\khp)$.
By analogy to Eq.~\eqref{eq:TqTqbar} we define
\begin{equation}
R_q^{\mu\nu}(\khp) \equiv \sum_{\beta=3}^5 R_\beta^{\mu \nu}(\khp) \, , \qquad R_{\qbar}^{\mu\nu}(\khp) \equiv \sum_{\beta=6}^8 R_\beta^{\mu \nu}(\khp)~,
\end{equation}
where $T_q^\mu(\khp)$ and $T_{\qbar}^\mu(\khp)$ can be obtained as $T_q^\mu(\khp) = R_q^{\mu -}(\khp)$ and $T_{\qbar}^\mu(\khp) = R_{\qbar}^{\mu -}(\khp)$.

Let us turn to contributions from diagrams (9)--(12) in
Fig.~\ref{fig:D2}.  Here, we look specifically at the diagram (9) that
yields,
\begin{equation}
  \begin{split}
  \calM_9^\mu(\MDEP) &= \int\frac{\rmd^4 k}{(2\pi)^4}
    \frac{\rmd^4 k_1}{(2\pi)^4}\ubar(-\rmi q_f e\gamma^\mu) S_0(q+k_\gamma)\calT(k,q+k_\gamma) S_0(q+k_\gamma-k)\\ 
&\times (-\rmi g\slashed{A}_{\rm LC}(k_1)\cdot t) \, S_0(q+k_\gamma-k-k_1)\calT(P - k - k_1,q+k_\gamma - k - k_1)\vp\,.
  \end{split}
\label{eq:m9lc}
\end{equation}
The integrations over $k^+$ and $k_1^+$ are trivial thanks to the nuclear effective vertex $\calT$. For the integrations over $k^-$ and $k_1^-$ we pick up the singularities in the quark propagators $S_0(q+k_\gamma-k)$ and in $S_0(q+k_\gamma-k-k_1)$.
We find
\begin{equation}
  \begin{split}
\calM_9^\mu(\MDEP)  &= q_f e g^2\int_{\kp\khp}\int_{\xp\yp}
    \frac{\rhop^a(\khp)}{\khp^2}\,
    \rme^{\rmi\kp\cdot\xp+\rmi(\Pp-\kp-\khp)\cdot\yp} \\
    &\quad\times \ubar R_9^{\mu i}(\kp,\khp)\frac{k_{1i}}{P^+}\bigl[\Uf(\xp)-1\bigr]
    t^a\bigl[\Uf^\dag(\yp)-1\bigr]\vp\,,
  \end{split}
\label{eq:M9}
\end{equation}
where
\begin{equation}
  R_9^{\mu\nu}(\kp,\khp) = \gmu \qkpropD\gp(\slashed{q}+\slashed{k}_\gamma-\slashed{k}+m)
    \gnu
    \frac{(\slashed{q}+\slashed{k}_\gamma-\slashed{k}-\slashed{k}_1+m)\gp}{N_k(\kp,\khp)}\,.
\label{eq:r9}
\end{equation}
A similar calculation can be performed for the remaining diagrams (10)--(12).
For $\beta=9, \dots, 12$, the difference from the Lorenz gauge calculation is to again
have the replacement $R_\beta^{\mu i}(k_{1 i}/P^+)$ instead of $-R_\beta^{\mu -}$. 
By analogy to Eq.~\eqref{eq:Tqqbar} we define
\begin{equation}
R_{q\qbar}^{\mu\nu}(\kp,\khp) \equiv \sum_{\beta=9}^{12} R_\beta^{\mu\nu}(\kp,\khp)\,.
\label{eq:rqq}
\end{equation}
Obviously, we can express $T_{q\qbar}^\mu(\kp,\khp)$ through $R_{q\qbar}^{\mu\nu}(\kp,\khp)$ as
\begin{equation}
  T_{q\qbar}^\mu(\kp,\khp) =
  R_{q\qbar}^{\mu -}(\kp,\khp)\,.
\label{eq:tqq_R}
\end{equation}
Finally, below we list the singularities we picked up when calculating the integrations over $k_1^-$ in the amplitudes with $\beta=9,\dots,12$
\begin{equation}
\begin{split}
& k^-_{1(9)} = \frac{M^2(\qp+\kgp-\kp)}{2(q^+ + k_\gamma^+)} + \frac{M^2(\qp+\kgp-\kp-\khp)}{2p^+} - \rmi\varepsilon\,,\\
& k^-_{1(10)} = \frac{M^2(\qp-\kp)}{2q^+} + \frac{M^2(\qp-\kp-\khp)}{2(p^+ + k_\gamma^+)} - \rmi\varepsilon\,,\\
& k^-_{1(11)} = \rmi \varepsilon\,, \qquad
k^-_{1(12)} = \rmi \varepsilon\,.
\end{split}
\label{eq:k1m9to12}
\end{equation}

By using the relations
\begin{equation}
\begin{split}
&  R_g^{\mu\nu}C_\nu(P,\Pp)  + \left[R_q^{\mu i}(\Pp)+R_{\qbar}^{\mu i}(\Pp)-R_{q\qbar}^{\mu i}(\boldsymbol{0},\Pp)\right]\frac{P_i}{P^+} = 0\,,\\
& R_q^{\mu i}(\Pp - \kp) - R_{q\qbar}^{\mu\nu}(\kp,\Pp-\kp) = 0 \,,\qquad R_{\qbar}^{\mu\nu}(\khp) - R_{q\qbar}^{\mu\nu}(\boldsymbol{0},\khp) = 0\,,
\end{split}
\label{eq:cancLC}
\end{equation}
we find that the terms in the amplitude that do not contain a Wilson line or contain only one Wilson line cancel. Eqs.~\eqref{eq:cancLC} are analogous to Eqs.~\eqref{eq:amp-ident}.
Now, the complete expression for the amplitude is thus,
\begin{equation}
  \begin{split}
  &\calM^\mu(\MDEP) = -q_f e g^2 \int_{\kp\khp}\int_{\xp\yp}
  \frac{\rhop^a(\khp)}{\khp^2}\,
  \rme^{\rmi\kp\cdot\xp+\rmi(\Pp-\kp-\khp)\cdot\yp} \\
  &\qquad\times \ubar \bigl\{ R_g^{\mu\nu}C_\nu(P,\khp)
  U^{ba}(\xp)t^b - R_{q\qbar}^{\mu i}(\kp,\khp)\frac{k_{1i}}{P^+}
  \Uf(\xp) t^a \Uf^\dag(\yp)\bigr\}\vp\,.
\end{split}
\label{eq:ampt}
\end{equation}

\subsection{Equivalence with the amplitude in the Lorenz gauge}

There are two apparent differences between the light-cone gauge amplitude in
Eq.~\eqref{eq:ampt} and the Lorenz gauge amplitude in Eq.~\eqref{eq:full-amp};  first,
$C^\nu(P,\khp)$ is replaced with the Lipatov vertex $C^\nu_L(P,\khp)$,
and second, the gluon vertex in the quantity
$T_{q\qbar}^\mu(\kp,\khp)$ carries a light-cone index instead of the
transverse index, as seen in Eq.~\eqref{eq:ampt}.  
We can show that
they are identical by two-step arguments.  First, let us look at the
difference between $C^\nu$ and $C^\nu_L$, that is,
\begin{equation}
  C^\nu(P,\khp) - C^\nu_L(P,\khp)
  = \biggl( \frac{\khp^2}{P^+ P^-} - 1\biggr) P^\nu
  + \frac{P^2}{P^+} \delta^{\nu +}\,.
\end{equation} 
We contract this difference with $R_g^{\mu\nu}$ and sandwich
it with the $\ubar$ and $\vp$ spinors. We find
\begin{equation}
\begin{split}
 \ubar R_g^{\mu\nu}\bigl[ C_{L\,\nu}(P,\khp) &-C_\nu(P,\khp) \bigr]\vp = \ubar \bigl[ R_g^{\mu\nu}C_{L\,\nu}(P,\khp)-T^\mu_g(\khp) \bigr]\vp\\
  &= -\frac{1}{P^+}\ubar \bigl[ \gmu \qkpropD\gp
 - \gp \pkpropD\gmu \bigr] \vp\,.
\end{split}
\label{eq:gi1}
\end{equation}
In the first line we used \eqref{eq:tglc} and in the second line we used the Dirac equation.

Next, we work on the part of the amplitude containing
$R_{q\qbar}^{\mu\nu}(\kp,\khp)$.
With some calculations we can prove that for $\beta=9,\dots,12$,
\begin{equation}
  \sum_{\beta=9}^{12}\ubar R_\beta^{\mu\nu}(\kp,\khp) \tilde{k}_{1(\beta) \nu} \vp = 0\,,
\label{eq:wip}
\end{equation}
where the four vector $\tilde{k}_{1(\beta)}$ is defined with
$\tilde{k}_{1(\beta)}^+ = P^+$ and $\tilde{k}_{1(\beta)}^-$ from the
singularity of the integrand, see Eqs.~\eqref{eq:k1m9to12}, as well as
$\tilde{\bk}_{1(\beta)\perp}=\khp$.

For $R_{11,12}^{\mu\nu}$ the singularity is located at
$k_{1(11)}^-=k_{1(12)}^-=0$, see Eq.~\eqref{eq:k1m9to12}, and so from Eq.~\eqref{eq:wip} we have,
\begin{align}
  &\ubar\bigl[ T_{q\qbar}^\mu(\kp,\khp)P^+
    + R_{q\qbar}^{\mu i}(\kp,\khp)k_{1i}\bigr] \vp \notag\\
  &\quad = -\ubar R_{q\qbar}^{\mu +}(\kp,\khp)k_{1(\beta)}^- \vp
   = -\ubar\bigl[ R_9^{\mu +}(\kp,\khp) k_{1(9)}^-
    + R_{10}^{\mu +}(\kp,\khp) k_{1(10)}^- \bigr]\vp\,,
\end{align}
where in the first line we used Eq.~\eqref{eq:tqq_R}.
We consider the first term for the moment and use the explicit expression
for $R^{\mu +}_9(\kp,\khp)$ from Eq.~\eqref{eq:r9} as well as the explicit form of $k_{1(9)}^-$, see Eq.~\eqref{eq:k1m9to12}. We then find,
\begin{align}
  \ubar R^{\mu +}_9(\kp,\khp) k_{1(9)}^- \vp
  & = \ubar\gmu \qkpropD\gp\frac{(\slashed{q}+\slashed{k}_\gamma - \slashed{k}+m)\gamma^+(\slashed{q}+\slashed{k}_\gamma - \slashed{k}-\slashed{k}_1+m) \vp}{-2p^+ \, 2(q^+ + k_\gamma^+)} \notag\\
  & = -\ubar \gmu \qkpropD\gamma^+ \vp\,.
\label{eq:R9alg}
\end{align} 
In the second line we moved
$\gp$ and used $(\gp)^2 = 0$.  The second term, containing $R_{10}^{\mu +}(\kp,\khp)$, can be manipulated in a
similar way, leading to
\begin{equation}
  \ubar\biggl[ T_{q\qbar}^\mu(\kp,\khp) + R_{q\qbar}^{\mu i}(\kp,\khp)
    \frac{k_{1i}}{P^+}\biggr] \vp
  = \frac{1}{P^+}\ubar\left[ \gmu \qkpropD\gp
    - \gp \pkpropD\gmu \right] \vp\,.
\label{eq:gi2}
\end{equation} 
This has exactly the same form as Eq.~\eqref{eq:gi1} but with an
opposite sign.  To end the proof we note that Eq.~\eqref{eq:gi2} is
independent of $\kp$ and so we effectively have $\xp = \yp$ in the
fundamental Wilson lines in the amplitude \eqref{eq:ampt}.  Then we can use the
following identity,
\begin{equation}
  \Uf(\xp)t^a \Uf^\dag(\xp) = t^b U^{ba}(\xp)\,,
\end{equation}
which will ensure that Eqs.~\eqref{eq:gi1} and \eqref{eq:gi2} cancel
out once these expressions are multiplied by the appropriate Wilson
lines.

\section{Properties of the photon production amplitude}
\label{app:Properties}

In this Appendix we will prove that the amplitude satisfies the photon Ward identity and the famous Low-Burnett-Kroll soft photon theorem \cite{Low:1958sn,Burnett:1967km,Bell:1969yw}. 
The Low-Burnett-Kroll theorem states that, in the infrared limit for the radiated photon momentum, i.~e. $k_\gamma\to 0$, the amplitude should factorize into the product of the non-radiative amplitude, the photon polarization vector, and a vectorial structure that depends on momenta of the emitted charged particles.

We will rely on diagrammatic representations in the derivation in Sec.~\ref{sec:AL} though our conclusions will be valid of course for another derivation in \ref{app:LC}.
 
\subsection{Photon Ward identity}

The photon Ward identity implies that the amplitude vector given in Eq.~\eqref{eq:full-amp} should satisfy
\begin{equation}
 k_{\gamma\mu} \calM^\mu(\MDEP)=0\,.
\label{eq:photonWI}
\end{equation}
As we demonstrate below, this is satisfied independently for $T^\mu_g$ and $T^\mu_{q\qbar}$ terms that constitute the total amplitude.  For $T_g^\mu$,
we immediately notice,
\begin{equation}
  \ubar\, k_{\gamma\mu} T^\mu_g(\khp)\,\vp
  = \frac{1}{P^2}\ubar\left[ \Lip(P,\khp)\frac{\slashed{p}-m}{2p\cdot k_\gamma}\slashed{k}_\gamma - \slashed{k}_\gamma\frac{\slashed{q}+m}{2q\cdot k_\gamma} \Lip(P,\khp) \right]\vp=0\,,
\end{equation}
which we can easily prove using $(\slashed{p}-m)\slashed{k}_\gamma=-\slashed{k}_\gamma(\slashed{p}-m)+2p\cdot k_\gamma$ in the first term and $\slashed{k}_\gamma(\slashed{q}+m)=-(\slashed{q}+m)\slashed{k}_\gamma+2q\cdot k_\gamma$ in the second term and the Dirac equations satisfied by $\ubar$ and $\vp$.
For $T^\mu_{q\qbar}$, it is somewhat more involved to prove a counterpart of the identity.  By definition as given in Eq.~\eqref{eq:Tqqbar} $T_{q\qbar}^\mu$ is a sum of $R_\beta^\mu$ with $\beta=9,\cdots,12$.  Using the explicit forms of $R_\beta^\mu$ in Eq.~\eqref{eq:MQ-beta}, we can prove the following relation,
\begin{equation}
  \ubar k_{\gamma\mu} \bigl[ R_{9}^\mu(\kp,\khp)+R_{11}^\mu(\kp,\khp)\bigr] \vp \notag\\
 =-\ubar k_{\gamma\mu} \bigl[ R_{10}^\mu(\kp,\khp)+R_{12}^\mu(\kp,\khp)\bigr] \vp\,.
\end{equation}
The different denominators in the expressions above, $N_k(\kp,\khp)$ in $R_9^\mu$ and $R_{11}^\mu$ and $N_q(\kp,\khp)$ in $R_{10}^\mu$ and $R_{12}^\mu$, cancel with the numerator after taking the contraction with the photon momentum, $k_{\gamma\mu}$.
This cancellation occurs in a way similar to the $T_g^\mu$ case as a consequence of anticommuting the gamma matrices and using the Dirac equations satisfied by $\ubar$ and $\vp$ as well as using the on-shell-ness of the photon momentum.
This leads to
\begin{equation}
  \ubar\, k_{\gamma\mu} T_{q\qbar}^\mu(\kp,\khp)\vp=0\,.
\end{equation}
Since the $T_g^\mu$ and the $T_{q\qbar}^\mu$ contributions separately vanish, we have confirmed that the photon Ward identity~\eqref{eq:photonWI} is certainly satisfied.

\subsection{Soft-photon factorization}
As we will demonstrate explicitly, the photon production amplitude we have derived satisfies the Low-Burnett-Kroll theorem: we will recover the non-radiative amplitude (and sub-leading pieces coming from the diagrams (11) and (12) in Fig.~\ref{fig:D2} in which the photon is not radiated from external legs)
The leading contributions encompass sub-processes in which the photon is radiated after the $q\qbar$ pair scatters off the nucleus, and thus the photon is attached to exteral legs. Such leading terms possess the factor,
\begin{equation}
\frac{\qks+m}{\qkprop} \qquad\qquad\text{or} \qquad\qquad \frac{\pks-m}{\pkprop}~.
\label{eq:quarkprop}
\end{equation} 
The numerators are finite in the $k_\gamma\to 0$ limit, and non-vanishing contributions under this limit are
\begin{align}
  \gmu(\pks-m)\vp   &\;\to\; (\slashed{p}-m)\gmu\vp = 2p^\mu \vp\,,\notag\\
  \ubar\gmu(\qks+m) &\;\to\; \ubar\gmu(\slashed{q}+m) = 2q^\mu \ubar\,,
\end{align}
while the denominators are linearly divergent in the $k_\gamma\to 0$
\begin{equation}
  \pk^2 - m^2 \;\to\; 2p\cdot k_\gamma\,,\qquad
  \qk^2 - m^2 \;\to\; 2q\cdot k_\gamma\,.
\end{equation}
With these simplifications, we find,
\begin{align}
  \epsilon_\mu^\ast(\kg,\lambda) T_g^\mu(\khp)
  &= \epsilon_\mu^\ast(\kg,\lambda)\frac{1}{P^2}\biggl[
    \Lip(P,\khp)\pkpropD\gmu-\gmu\qkpropD\Lip(P,\khp) \biggr] \notag\\
  &\to \epsilon_\mu^\ast(\kg,\lambda)\biggl(
  \frac{p^\mu}{p\cdot k_\gamma}
  -\frac{q^\mu}{q\cdot k_\gamma}\biggr)T_g(\khp)\,,
\end{align}
where the last piece defined by
\begin{equation}
  T_g(\khp) \equiv \frac{\Lip(p+q,\khp)}{(p+q)^2}\,,
\end{equation}
is the expression obtained in the computation of the amplitude for quark-antiquark pair production~\cite{Blaizot:2004wv}.

For $T^\mu_{q\qbar}$, at the leading order, we only have to concern ourselves with diagrams (9) and (10) in Fig.~\ref{fig:D2}. This is because the contributions from (11) and (12) do not contain any quark propagator in the form of Eq.~\eqref{eq:quarkprop}, while (9) and (10) are divergent in the $k_\gamma\to 0$ limit, and this explains our identification of (9) and (10) as leading contributions and (11) and (12) as sub-leading ones. From this argument we can show that the $T_{q\qbar}^\mu$ term with leading contributions only behaves as follows,
\begin{equation}
  \epsilon_\mu^\ast(\kg,\lambda) T^\mu_{q\qbar}(\kp,\khp)
  \;\to\; \epsilon_\mu^\ast(\kg,\lambda) \biggl(
  \frac{p^\mu}{p\cdot k_\gamma}
  -\frac{q^\mu}{q\cdot k_\gamma}\biggr)
  T_{q\qbar}(\kp,\khp)\,,
\end{equation}
where the last piece defined by
\begin{equation}
  T_{q\qbar}(\kp,\khp) \equiv
  \frac{\gp(\qps-\kps+m)\gm(\qps-\kps-\khps+m)\gp}
       {2p^+M^2(\qp-\kp)+2q^+M^2(\qp-\kp-\khp)}\,,
\end{equation}
is the expression that also appears as a part of the amplitude for quark-antiquark pair production~\cite{Blaizot:2004wv}. It should be noted that in the computation in Ref.~\cite{Blaizot:2004wv} only the leading terms have been concerned, while the $k_\gamma\to 0$ limit of the amplitude in the present work can also pick up the sub-leading rest from diagrams (11) and (12).
Combining both sets of leading contributions in the soft photon limit, we then get,
\begin{equation}
  \epsilon_\mu^\ast(\kg,\lambda)\calM^\mu(\MDEP)
  \;\to\; -q_f e\,\epsilon_\mu^\ast(\kg,\lambda)\biggl(
  \frac{p^\mu}{p\cdot k_\gamma}
  -\frac{q^\mu}{q\cdot k_\gamma}\biggr)\calM(\bq,\bp)\,,
\label{eq:LBK}
\end{equation} 
where $\calM(\bq,\bp)$ is the amplitude for $q\qbar$ pair production found in Ref. \cite{Blaizot:2004wv}, and Eq.~\eqref{eq:LBK} verifies that the Low-Burnett-Kroll theorem in the $k_\gamma\to 0$ limit is satisfied.

\bibliographystyle{elsarticle-num}
\bibliography{references}{}

\end{document}